\documentclass[prb,twocolumn,showpacs,amsmath,amssymb,superscriptaddress]{revtex4}
\usepackage{mathrsfs}
\usepackage{graphicx}
\usepackage{bm}
\usepackage{amsmath}
\usepackage{multirow}
\usepackage{xcolor}
\usepackage{txfonts}
\begin{document}

\title{Unconventional superconductivity in bilayer transition metal dichalcogenides}
\author{Chao-Xing Liu} 
\affiliation{$^{1}$Department of Physics, The Pennsylvania State University, University Park, Pennsylvania 16802-6300, USA;}

\begin{abstract}
Bilayer transition metal dichalcogenides (TMDs) belong to a class of materials with two unique features,
the coupled spin-valley-layer degrees of freedom and the crystal structure that is globally centrosymmetric but locally non-centrosymmetric.
In this work, we will show that the combination of these two features can lead to a rich phase diagram for unconventional superconductivity, including intra-layer and inter-layer singlet pairings
and inter-layer triplet pairings, in bilayer superconducting TMDs. In particular, we predict that the inhomogeneous Fulde-Ferrell-Larkin-Ovchinnikov state can exist in bilayer TMDs under an in-plane magnetic field.
We also discuss the experimental relevance of our results and possible experimental signatures.
\end{abstract}

\date{\today}

\maketitle

{\it Introduction.}--
Unconventional superconductivity \cite{sigrist1991a,mineev1999a,bauer2012}, which is beyond the simple s-wave spin-singlet superconductivity in the Bardeen-Cooper-Schrieffer theory, can emerge in two dimensional (2D) systems, such as surfaces \cite{dimitrova2003a,agterberg2003a,barzykin2002} or interfaces \cite{aoyama2012}, superconducting heterostructures \cite{yoshida2013} and 2D or quasi-2D superconducting materials \cite{houzet2002,yoshida2012,lu2015,xi2015,saito2016,navarro2016}. Recently, it was demonstrated that "Ising" superconductivity can exist in monolayer transition metal dichalcogenides (TMDs), such as MoS$_2$ \cite{saito2016,lu2015} and NbSe$_2$ \cite{xi2015}, based on experimental observation that in-plane upper critical field $H_{c2,\parallel}$ is far beyond the paramagnetic limit. The space symmetry group of the monolayer TMD is the $D_{3h}$ group without inversion symmetry. Thus, the monolayer superconducting TMDs belong to the so-called non-centrosymmetric superconductors (SCs) \cite{bauer2012}, for which spin-up and spin-down Fermi surfaces are split by strong spin-orbit coupling (SOC), leading to a mixing of spin singlet and triplet pairings \cite{zhou2016a,yuan2014}. 
The existence of triplet components can enhance
$H_{c2,\parallel}$ in non-centrosymmetric SCs \cite{frigeri2004}.
In monolayer TMDs, Ising SOC fixes spin axis along the out-of-plane direction
and greatly reduces the Zeeman effect of in-plane magnetic fields,
thus explaining the experimental observations of high $H_{c2,\parallel}$.
A high $H_{c2,\parallel}$ was also observed in bilayer TMDs (e.g. NbSe$_2$) \cite{xi2015}. The crystal structure of bilayer TMDs is described by the symmetry group $D_{3d}$ with inversion symmetry and the corresponding Fermi surfaces are spin degenerate. This experimental result motivates us to study the difference between bilayer superconducting TMDs and conventional SCs.

We first illustrate the difference from symmetry aspect. Although inversion symmetry exists in bilayer TMDs, the inversion center should be chosen at the center between two layers, labeled by "P" in Fig. 1a. As a result, bilayer TMDs belong to a class of materials which are globally centrosymmetric, but locally non-centrosymmetric (for each layer). The absence of local inversion symmetry can lead to the "hidden" spin polarization \cite{zhang2014a,riley2014}, the spin-layer locking \cite{dong2015, jones2014} and other exotic physical phenomena \cite{liu2013a}. The superconductivity for these materials has been studied in the CeCoIn$_5$/YbCoIn$_5$ hybrid system \cite{sigrist2014,yoshida2012}, SrPtAs \cite{goryo2012,fischer2011,sigrist2014,youn2012} and other bilayer Rashba systems \cite{nakosai2012}. Inhomogeneous Fulde-Ferrell-Larkin-Ovchinikov (FFLO) states were proposed in CeCoIn$_5$/YbCoIn$_5$ hybrid system while chiral topological $d+id$ superconductivity was suggested in SrPtAs. Bilayer TMDs possess global $D_{3d}$ symmetry and local $D_{3h}$ symmetry, labeled as $D_{3d}(D_{3h})$, and thus it is equivalent to that of SrPtAs \cite{fischer2011}, but different from CeCoIn$_5$/YbCoIn$_5$ hybrid system with $D_{3d}(C_{3v})$ symmetry. Due to the $D_{3h}$ symmetry in each layer, Ising SOC is expected in bilayer TMDs and SrPtAs, while Rashba SOC occurs in CeCoIn$_5$/YbCoIn$_5$ hybrid system.

In this work, we study possible superconducting pairings 
based on a prototype model of bilayer TMDs.
The superconducting phase diagram as a function intra-layer ($U_0$) and inter-layer ($V_0$) interactions
is summarized in Fig. 1c, in which
three different pairings, intra-layer $A_{1g}$ pairing, intra-layer $A_{1u}$ pairing and inter-layer $E_u$ pairing,
can exist,
depending on the strength and sign of $U_0$ and $V_0$.
We further study the stability of these superconducting pairings under external magnetic fields.
In particular, we predict 
the FFLO state with a finite momentum pairing \cite{larkin1965,fulde1964}
induced by the orbital effect of in-plane magnetic fields.

\begin{figure} 
\includegraphics[width=8.6cm]{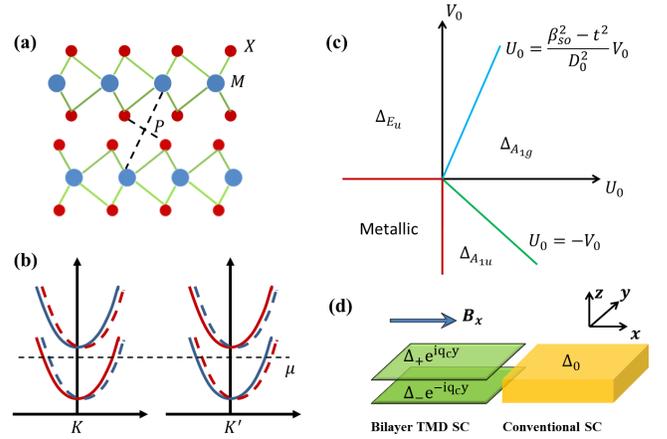}
\caption{(a) Crystal structure of bilayer TMDs $MX_2$ with the inversion center labelled by $P$.
(b) Schematics for energy dispersion of bilayer TMDs where red and blue are for spin up and spin down,
and solid and dashed lines are for the top and bottom layers. Here each band is doubly degenerate
and we shift the dashed lines a little for the view. (c) The phase diagram as a function of $U_0$ and $V_0$.
 The red, blue and green lines are the phase boundary, separating three superconducting phases, the $A_{1g}$, $A_{1u}$ and $E_u$ pairings, and the metallic phase. (d) Experimental setup of bilayer TMD SC/conventional SC junction. }   \label{Fig:1}
\end{figure}

{\it Phase diagram of bilayer TMDs} --
A prototype model for TMDs \cite{zhou2016a,yuan2014,xiao2012} was first derived for the conduction band of MoS$_2$ and can also be applied to other TMDs. This model is constructed on a triangle lattice of Mo atoms with $4d_{z^2}$ orbitals for each monolayer. The conduction band minima appear at two momenta $\pm K$, and one can regard $\pm K$ as valley index and expand the tight-binding model around $\pm K$ for each layer, as described in Ref. \cite{zhou2016a,yuan2014}. We extend this model to bilayer TMDs by including layer index. Let us label the annihilation fermion operator as $c_{\sigma,\eta}$, where $\sigma=\uparrow,\downarrow$ is for spin and $\eta=\pm$ is for two layers. On the basis of $(c_{\uparrow,+},c_{\downarrow,+},c_{\uparrow,-},c_{\downarrow,-})$, the effective Hamiltonian is
\begin{eqnarray}
\hat{H}_0({\bf p=\epsilon K+k})=\xi_{k}+\epsilon\beta_{so}s_z\tau_z+t\tau_x
\label{Eqn:H0}
\end{eqnarray}
where $s$ and $\tau$ are two sets of Pauli matrices for spin and layer degrees,
$\epsilon=\pm$ is for valley index and $\xi_{k}=\frac{\hbar^2}{2m}k^2-\mu$
with chemical potential $\mu$.
Here the $\beta_{so}$ term is the Ising SOC while the $t$ term is the hybridization between two layers.
The eigen-energy is given by $\varepsilon_{s,\lambda}=\xi_k+\lambda D_0$ with 
$D_0=\sqrt{\beta_{so}^2+t^2}$ and $s,\lambda=\pm$.
$s$ does not appear and thus the eigen-states with opposite $s$ are degenerate, as
shown in Fig. 1b.
We next consider the symmetry classification of superconducting pairings, 
similar to that in Cu doped Bi$_2$Se$_3$ SCs \cite{fu2010}
since both materials belong to $D_{3d}$ group.
We only consider s-wave pairing, and thus the gap function $\hat{\Delta}$ is
independent of momentum and can be expanded in terms of $s$ and
$\tau$ ($\hat{\Delta}=\sum_{i,\mu}\Delta_{i,\mu}\gamma_{i,\mu}$ where
$\gamma_{i,\mu}$ is a $4\times4$ matrix composed of $s$ and $\tau$ and $i,\mu$
are the indices labelling different representations).
Due to anti-commutation relation between fermion operators, the gap function needs to be anti-symmetric, and thus only six matrices
$s_y,s_y\tau_x,s_y\tau_z,\tau_y,s_x\tau_y,s_z\tau_y$ can couple to s-wave pairing.
The classification of these representation matrices, as well as their explicit physical meanings,
are listed in the Table I, from which $\Delta_{A_{1g},1}$ and $\Delta_{A_{1u}}$
describe intra-layer singlet pairings, $\Delta_{A_{1g},2}$ and $\Delta_{A_{2u}}$ give inter-layer singlet pairings
while $\Delta_{E_u,1}$ and $\Delta_{E_u,2}$ are inter-layer triplet pairings.
The pairing interaction can also be decomposed into different pairing channels as
$V_{A_{1g},1}=V_{A_{1u}}=\frac{U_0}{2}$ and
$V_{A_{1g},2}=V_{A_{2u}}=V_{E_u,1}=V_{E_u,2}=\frac{V_0}{2}$ (See appendix for details).

\begin{table}[htb]
  \centering
  \begin{minipage}[t]{1\linewidth}
	  \caption{ The matrix form and the explicit phyiscal meaning
	  of Cooper pairs in the representations $A_{1g}$, $A_{1u}$,
	  $A_{2u}$ and $E_u$ of the $D_{3d}$ group. Here $c_{\sigma\eta}$
	  is electron operator with  $\eta=\pm$ for layer index $\sigma$ for spin. 
	  $s$ and $\tau$ are Pauli matrices for spin and layer.
	  }
\hspace{-1cm}
\begin{tabular}
[c]{|c|c|c|}\hline\hline
Representation   &  Matrix form  & Explicit form \\ \hline
$A_{1g}$:
$\left.
\begin{array}{c}
       \Delta_{A_{1g},1}\\
       \Delta_{A_{1g},2}
\end{array}
\right.$  &
$\left.
\begin{array}{c}
	s_y\\
	s_y\tau_x
\end{array}
\right.$  & $\left.
\begin{array}{c}
	c_{\uparrow+}c_{\downarrow+}+c_{\uparrow-}c_{\downarrow-}\\
        c_{\uparrow+}c_{\downarrow-}+c_{\uparrow-}c_{\downarrow+}
\end{array}
\right. $ \\ \hline
$A_{1u}$: $\Delta_{A_{1u}}$
 & $s_y\tau_z$ &
 $c_{\uparrow+}c_{\downarrow+}-c_{\uparrow-}c_{\downarrow-}$
 \\ \hline
$A_{2u}$: $\Delta_{A_{2u}}$
 & $s_x\tau_y$ &
 $c_{\uparrow+}c_{\downarrow-}-c_{\uparrow-}c_{\downarrow+}$
 \\ \hline
$E_u$:
$\left.
\begin{array}{c}
       \Delta_{E_u,1}\\
       \Delta_{E_u,2}
\end{array}
\right.$  &
$\left.
\begin{array}{c}
	\tau_y\\
	s_z\tau_y
\end{array}
\right.$  & $\left.
\begin{array}{c}
	c_{\uparrow+}c_{\uparrow-}\\
        c_{\downarrow+}c_{\downarrow-}
\end{array}
\right. $ \\
\hline\hline
\end{tabular}
  \end{minipage}
\end{table}

Possible superconducting pairings are studied based on the linearized gap equations 
\cite{bauer2012,sigrist1991a,mineev1999a} (See appendix).
Around the valley $K$ (or $-K$), the Fermi surfaces for two spin states in each layer
are well separated by Ising SOC $\beta_{so}$ term.
Therefore, we below assume the Fermi energy only crosses the lower energy band at each valley
(Fig. 1b), for simplicity.
The pairings with different representations
do not couple to each other and thus,
we can compute the critical temperature $T_c$ in each representation, separately.
The critical temperature normally takes the form
$kT_{c0,i}=\frac{2\gamma\omega_D}{\pi} exp\left(- \frac{1}{N_0V_{i,eff}} \right)$, 
with the representation index $i$,
density of states $N_0$, the Debye frequency $\omega_D$
and $\gamma\approx 1.77 $. 
The effective interaction is given by $V_{A_{1g},eff}=2U_0+2V_0\frac{t^2}{D_0^2}$
for the $A_{1g}$ pairing, $V_{A_{1u},eff}=2U_0\frac{\beta^2_{so}}{D_0^2}$ for the $A_{1u}$ pairing 
and $V_{E_u,eff}=2V_0\frac{\beta^2_{so}}{D_0^2}$ for
the $E_u$ pairing, from which the corresponding critical temperature in each channel can be determined.
The $A_{2u}$ pairing does not exist because $V_{A_{2u},eff}=0$.
The phase diagram can be extracted by comparing different $T_{c0,i}$ (Fig. 1c).
The $A_{1g}$ pairing is favored by strong attractive intra-layer interaction ($U_0>0$), 
while the $E_{u}$ pairing
is favored by strong attractive inter-layer interaction ($V_0>0$). 
These two phases are separated by the critical line
$U_0=\frac{\beta_{so}^2-t^2}{D_0^2}V_0$. The $A_{1u}$ pairing appears
when the repulsive inter-layer interaction
is stronger than the attractive intra-layer interaction ($-V_0>U_0>0$) 
because repulsive inter-layer
interaction will favor opposite phases of pairing functions between two layers.
The $A_{1u}$ phase is separated from the $A_{1g}$ phase by a critical line
$U_0=-V_0$. When both $U_0$ and $V_0$ are repulsive interaction ($U_0,V_0<0$), no superconductivity can exist. 
For the 2D $E_u$ pairing, $\Delta_{E_u,1}$ and $\Delta_{E_u,2}$ are degenerate.
By taking into account the fourth order term in the Landau free energy (See Appendix), 
either nematic superconductivity $(\Delta_{E_u,1},\Delta_{E_u,2})=\Delta_{E_u}(\cos\theta,\sin\theta)$ ($\theta$ is a constant) \cite{fu2014} or chiral superconductivity with $(\Delta_{E_u,1},\Delta_{E_u,2})=\Delta_{E_u}(1,i)$ can be stabilized\cite{ueda1985}.

{\it Magnetic field effect} --
Next we study the effect of magnetic fields on bilayer superconducting TMDs.
Generally, magnetic fields have two effects, the Zeeman effect and the orbital effect.
The Zeeman coupling is given by
\begin{eqnarray}
	\hat{H}_{Zee}=g{\bf B}\cdot{\bf s}
\end{eqnarray}
where ${\bf B}$ labels the magnetic field and the Bohr magneton is absorbed into $g$ factor.
The orbital effect is normally included by replacing the momentum ${\bf k}$ in $\xi_k$ with
the canonical momentum ${\bf \pi}={\bf k}+\frac{e}{\hbar}{\bf A}$ with vector potential ${\bf A}$
(Peierls substitution). The orbital effect of in-plane magnetic fields is normally not important for a quasi-2D system.
However, it is not the case in bilayer TMDs due to its unusual band structure. Let's choose
${\bf A}=(0,-B_xz,0)$ for the in-plane magnetic field $B_x$, in which
the origin $z=0$ is located at the center between two layers.
As a result, $\xi_k$ is changed to
$\xi_{\pi}=\frac{\hbar^2}{2m}(k_x^2+(k_y-\frac{eB_xz_0}{2\hbar}\tau_z)^2)-\mu$ after the substitution, 
where $z_0$ is the distance between two layers.

The Ginzburg-Landau free energy is constructed as
\begin{eqnarray}
	L=\frac{1}{2}\sum_{ {\bf q},i\mu}\Delta^*_{i,\mu}({\bf q})\left(
	\frac{1}{V_i}\delta_{ij}\delta_{\mu\nu}-\chi^{(2)}_{ij,\mu\nu}({\bf q},{\bf B})
	\right)\Delta_{j,\nu}({\bf q})+L_4,
  \label{Eqn:LandauEnergy}
\end{eqnarray}
where $L_4$ describes the fourth order term. 
The superconductivity susceptibility $\chi^{(2)}_{ij,\mu\nu}$ can be expanded 
up to the second order of ${\bf q}$ and ${\bf B}$ ($q_iq_j$, $B_iB_j$ and
$q_iB_j$ with $i,j=x,y,z$). The magnetic field correction
to $T_{c0,i}$ for different pairings can be extracted by minimizing the above free energy
(See appendix). 

Due to the orbital effect, the Hamiltonian (\ref{Eqn:H0}) is changed to
\begin{eqnarray}
     \hat{H}'_0=\xi_k-\hbar v_Q k_y\tau_z+\epsilon \beta_{so}s_z\tau_z+t\tau_x, \label{Eqn:H01}	
\end{eqnarray}
where $v_Q=\frac{eB_xz_0}{2m}$ and the chemical potential $\mu$ in $\xi_k$ is re-defined to
include the $B_x^2$ term. We first focus on the limit $t\rightarrow 0$, in which
the energy dispersion of the Hamiltonian (\ref{Eqn:H01}) is shown in Fig. 2a.
The energy bands on the top and bottom layers are shifted in the opposite directions in the momentum
space by $Q=\frac{eB_xz_0}{2\hbar}$. This momentum shift cannot be ``gauged away'' and
thus the intra-layer spin-singlet pairing must carry a non-zero total momentum. 
This immediately suggests the possibility of
the FFLO state \cite{larkin1965,fulde1964,casalbuoni2004}
for the intra-layer singlet $A_{1g}$ and $A_{1u}$ pairings.
Since in-plane magnetic fields break the $D_{3d}$ symmetry, 
the orbital effect can mix the singlet $A_{1g}$ and $A_{1u}$ pairings. In the limit
$t\rightarrow 0$ with $T_{c0,A_{1g}}=T_{c0,A_{1u}}=T_{c0}$, we derive the 
free energy for the coupled $A_{1g}$ and $A_{1u}$ pairings as
\begin{eqnarray}
	&&L_2=\frac{1}{2}\sum_{\bf q}\left[\left( 4N_0 ln\left( \frac{T}{T_{c0}}\right)-\mathcal{P}(h_x,{\bf q})
	\right)\sum_{i=A_{1g},A_{1u}}|\Delta_i|^2\right.\nonumber\\
	&&\left. - \Delta_{A_{1g}}^*\mathcal{Q}\Delta_{A_{1u}} -\Delta_{A_{1u}}^* \mathcal{Q}\Delta_{A_{1g}}\right],
\end{eqnarray}
in which the detailed form of $\mathcal{P}$ and $\mathcal{Q}$ are defined in Appendix.
The term $\mathcal{Q}= \tilde{K} B_xq_y$ with a constant $\tilde{K}$ 
mixes $A_{1g}$ and $A_{1u}$ pairings.
With a transformation $\Delta_\pm=\frac{1}{\sqrt{2}}\left( \Delta_{A_{1g}}\pm\Delta_{A_{1u}} \right)$,
the free energy is changed to
\begin{eqnarray}
	L_2=\frac{1}{2}\sum_{\alpha=\pm, {\bf q}}\left( 4N_0 ln\left( \frac{T}{T_{c0}}\right)-\mathcal{P}(B_x,{\bf q})
	-\alpha \mathcal{Q}(B_x,{\bf q})\right)|\Delta_\alpha|^2.\label{Eqn:L2layer}
\end{eqnarray}
The corresponding critical temperature is determined by maximizing
$ln\left( \frac{T_c}{T_{c0}} \right)=\frac{1}{4N_0}\left( 
\mathcal{P}(B_x,{\bf q})+\alpha\mathcal{Q}(B_x,{\bf q}) \right)$ 
with respect to ${\bf q}$ and $\alpha$. 
From the explicit form of $\mathcal{P}$ and $\mathcal{Q}$,
the maximum is achieved by $q_x=0$ and 
$|q_y|=q_c=\frac{eB_xz_0}{\hbar}=2Q$,
thus realizing the FFLO state. 
The corresponding correction to $T_c$ vanishes ($T_{c}=T_{c0}$).
As a comparison, the $T_c$ of zero momentum pairing decreases with magnetic fields as
$ ln \left( \frac{T_c({\bf q}=0)}{T_{c0}}\right)=-C \left( \frac{\hbar v_Q k_f}{2\pi kT} \right)^2
\propto -B_x^2$
and the FFLO state is always favored in the limit $t\rightarrow 0$ for in-plane magnetic fields.

The form of the stable pairing function depends on the sign of $\mathcal{Q}$. 
Let's assume $B_x>0$ and $\tilde{K}>0$ in $\mathcal{Q}= \tilde{K} B_x q_y$. 
If $q_y=q_c>0$, $\mathcal{Q}>0$ and thus $\Delta_+$ pairing is favored. 
If $q_y=-q_c<0$, $\mathcal{Q}<0$ and $\Delta_-$ is favored.
$\Delta_+(q_c)$ and $\Delta_-(-q_c)$ are degenerate for the second order term of free energy. 
The FFLO state in the real space is
\begin{eqnarray}
	\Delta({\bf r})=\Delta_+(q_c)e^{iq_cy}+\Delta_-(-q_c)e^{-iq_cy}.\label{Eq:Delta_A1gA1u}
\end{eqnarray}
The exact form of pairing function is determined by the fourth order term 
of $\Delta_+(q_c)$ and $\Delta_-(-q_c)$, 
which is phenomenologically given by
\begin{eqnarray}
	L_4=\mathcal{B}_s\left( |\Delta_+(q_c)|^2+|\Delta_{-}(-q_c)|^2 \right)^2+
	\mathcal{B}_a\left( |\Delta_+(q_c)|^2 - |\Delta_-(-q_c)|^2\right)^2.\nonumber\\
\end{eqnarray}
If $\mathcal{B}_a>0$, we need $|\Delta_+(q_c)|=|\Delta_-(-q_c)|=\Delta_0$ to minimize $L_4$.
This state is known as LO phase \cite{larkin1965,houzet2001} or
stripe phase \cite{yoshida2013,dimitrova2003a,agterberg2007,barzykin2002}
or pair density wave \cite{yoshida2012,chen2004,soto2014}.
If $\mathcal{B}_a<0$, we have either $\Delta_+(q_c)=0$ or $\Delta_-(-q_c)=0$. 
In either case, the amplitude of 
$\Delta({\bf r})$ is fixed while its phase oscillates, thus correponding
to FF phase \cite{fulde1964,houzet2001} or helical phase 
\cite{agterberg2007,dimitrova2007,kaur2005,bauer2012,michaeli2012}. 
In the limit $t\rightarrow 0$, the coefficients are computed as
$\mathcal{B}_s=\mathcal{B}_{a}=\frac{7N_0\zeta(3)}{16(\pi kT_{c0})^2}>0$. 
Therefore, the stripe phase will be favored 
under an in-plane magnetic field near the critical temperature.

In the limit $t\rightarrow 0$,  
$\Delta_+$ and $\Delta_-$ are just the singlet pairing 
on the top and bottom layers according to Table I, 
and the free energies for $\Delta_+$ and $\Delta_-$ become decoupled  
(see Eq. (\ref{Eqn:L2layer}) for $L_2$ term and Eq. (96) of the appendix for $L_4$ term). 
Thus, the FFLO state in Eq. (\ref{Eq:Delta_A1gA1u})
can be viewed as two independent helical phases in two separate layers. 
No supercurrent or other observables can exist in helical phases
\cite{dimitrova2007,kaur2005} for infinite large systems. 
To identify this phase, one needs to consider a Josephson
junction structure between bilayer TMDs and conventional SCs (Fig. 1d), 
similar to that discussed in Ref. \cite{kaur2005,bauer2012,yang2000}
(See appendix for details). 
For a finite tunneling $t$, the interference between two layers leads to 
the gap oscillation of stripe phase in Eq. (\ref{Eq:Delta_A1gA1u}).  

We notice that the FFLO phase has been proposed in non-centrosymmetric SCs under a magnetic field
\cite{barzykin2002,sigrist2014},
and emphasize two essential differences between our case and non-centrosymmetric SCs.
(1) In non-centrosymmetric SCs, the FFLO phase is induced by a linear gradient term
$\tilde{K}_{ij}\Delta^*B_iq_j\Delta$ ($\tilde{K}_{ij}$ is a parameter) that breaks inversion
symmetry. In contrast, inversion symmetry is preserved in our system, and the linear gradient
term ($\tilde{K}_{ij}\Delta_{A_{1g}}^*B_iq_j\Delta_{A_{1u}}$)
couples two pairings with opposite parities.
(2) In non-centrosymmetric SCs, the FFLO phase results from the combination of
Rashba SOC and Zeeman effect of magnetic fields.
In our system, the FFLO phase is from the combination of Ising SOC 
and the orbital effect
of magnetic fields. In particular, this phase can occur for any magnetic field strength in the weak
interlayer coupling limit $t\rightarrow0$.

When $t\neq 0$, the occurence of the FFLO phase will be shifted to a finite magnetic field.
We numerically minimize free energy with respect to the momentum ${\bf q}$
and calculate the magnetic field correction to $T_c$.  
In Fig. 2b, $T_c/T_{c0}$
is plotted as a function of magnetic field $B_x$ for three hybridization parameters $t$. 
The momenta for the corresponding stable states, labeled by $q_c$, are shown in Fig. 2c.
For a weak hybridization ($t=1$meV $\ll \beta_{so}=40$meV),
FFLO phase appears at a small $B_x$,
and the corresponding $q_c$ approaches $2Q$ with increasing $B_x$.
There is only a weak correction to $T_c$ for the FFLO phase (black line in Fig. 2b).
When increasing hybridization ($t=5, 10$meV), 
zero momentum pairing is favored for small $B_x$
and lead to a rapid decrease of $T_c$ with its correction
given by $\frac{T_c-T_{c0}}{T_{c0}}\propto -B_x^2$ (red and blue lines in Fig. 2b).
When $B_x$ becomes larger, a transition from zero momentum 
pairing to the FFLO state occurs. The decreasing
in $T_c$ deviates from the $B_x^2$ dependence and becomes weaker. 
Experimentally, one can control
the hybridization between two layers by inserting an insulating layer in between,
and the deviation of the $T_c$ correction from the $B_x^2$ dependence
implies the occurrence of FFLO states in this system.
We further construct the phase diagram by evaluating gap functions
as a function of temperatures and magnetic fields
for $t=10meV$ in Fig. 2d. As discussed in appendix, 
The transition from the normal metal (III region in
Fig. 2d) to uniform SC (I region) or FFLO state (II region) is 
of the second order type (dashed red line in Fig. 2d) 
while the transition between uniform SC and FFLO state is 
of the first order type (dashed black line in Fig. 2d).

Besides the orbital effect, the correction of $T_c$ due to the Zeeman effect, which
is the same for zero-momentum pairing and the FFLO phase, is given by
$ln \left(\frac{T_{c,A_{1g}}}{T_{c0,A_{1g}}}
\right)\propto -\frac{t^2}{\beta_{so}^2}B_{x}^2$ for $A_{1g}$ pairing
and $ln \left(\frac{T_{c,A_{1g}}}{T_{c0,A_{1g}}}\right)\propto -\frac{t^4}{\beta_{so}^4}B_{x}^2$
for $A_{1u}$ pairing. Additional factors $t^2/\beta_{so}^2$ and
$t^4/\beta_{so}^4$ greatly reduce the $B_x^2$ dependence for the $A_{1g}$ and $A_{1u}$
pairings in the limit $t\ll \beta_{so}$. The behavior of out-of-plane magnetic field ($B_z$)
in bilayer TMDs is similar to that of conventional SCs (See Appendix).

\begin{figure} 
\includegraphics[width=8.6cm]{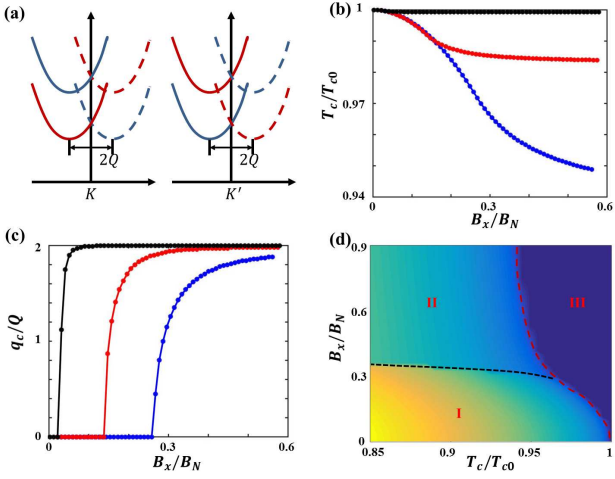}
\caption{(a) Schematics of energy dispersion for bilayer TMDs 
with an in-plane magnetic field. Here red and blue colors are for opposite spins
and solid and dashed lines are for top and bottom layers. 
(b) The magnetic field dependence of the critical temperature
$T_c$. Here the black line is for $t=1meV$, 
the red is for $t=5meV$ while the blue is for $t=10meV$.
Other parameters are chosen as $\beta_{so}=40meV$, 
$\hbar v_F= 30 meV\cdot nm$ and $m=0.6 m_e$ with
electron mass $m_e$, $N_0U_0=0.3$ and $N_0V_0=0.1$. 
Only the orbital effect is taken into account. 
(c) The momentum $q_c$ for the stable pairing state as a function of 
$B_x$. (d) Phase diagram as a function of $B_x$ and $T_c$. 
Here I is for conventional SC phase, II is for FFLO state
and III is for normal metal. $B_N=\frac{2kT_{c0}}{v_fz_0}$. }
\label{Fig:2}
\end{figure}

{\it Discussion and Conclusion } --
In realistic bilayer superconducting TMDs, the Fermi energy will cross both spin states in each layer.
However, once the Ising SOC is larger than other energy scales ($\beta_{so}\gg t, \hbar k_f v_Q, \hbar v_f q$),
the Fermi surfaces for two spin states in one layer are well separated 
and the physics discussed here should be valid qualitatively.
Based on the existing experiments, the $A_{1g}$ pairing 
is mostly likely to exist at a zero magnetic field. In this case, we predict the occurence of
the FFLO phase 
under an in-plane magnetic field.
The onset magnetic field is determined by the ratio between
inter-layer hybridization $t$ and Ising SOC $\beta_{so}$
($\frac{t}{\beta_{so}}\sim 0.27$ in NbSe$_2$) \cite{xi2015}.
Our results suggest a weak correction to $T_c$ 
for both the orbital and Zeeman effects of in-plane magnetic fields,
thus consistent with experimental observations of 
high in-plane critical fields in
bilayer superconducting TMDs \cite{xi2015}.
The central physics in this work originates from the unique
crystal symmetry property,
and similar physics can occur in SrPtAs \cite{fischer2011}.
Similar physics also occurring for exciton condensate in a bilayer
system \cite{efimkin2011,seradjeh2012}.
Our work paves a new avenue to search for unconventional superconductivity in
2D centrosymmetric SCs.

%

{\it Acknowledgement}
We would like to thank Xin Liu, K. T. Law and Kin Fai Mak for the helpful discussion.
C.-X. Liu acknowledges the support from Office of Naval Research (Grant No. N00014-15-1-2675).

\appendix

\begin{widetext}

\section{Landau-Ginzburg free energy and linearized self-consistent gap equation}
In this section, we review the formalism for Landau-Ginzburg free energy and linearized
self-consistent gap equation \cite{dimitrova2007,mineev1999a,samokhin2004}, which will be used in the main text. We may start from
the interacting Hamiltonian
\begin{eqnarray}
	H=H_0+H_V=\sum_{ {\bf p,p'},\alpha\beta}\hat{c}^{\dag}_{ {\bf p}\alpha}(H_0)_{\alpha\beta}\hat{c}_{ {\bf p'},\beta}
	+\frac{1}{2}\sum_{ {\bf pp'q},\alpha\beta\gamma\delta}V_{\alpha\beta\gamma\delta}({\bf p,p',q})
	\hat{c}^{\dag}_{ {\bf \frac{q}{2}-p},\alpha}\hat{c}^{\dag}_{ {\bf \frac{q}{2}+p},\beta}
	\hat{c}_{ {\bf \frac{q}{2}+p'},\gamma}\hat{c}_{ {\bf \frac{q}{2}-p'},\delta},
\end{eqnarray}
where the $H_0$ term is for single-particle Hamiltonian and the $V$ term is for interaction.
We may consider the path integral formalism of the BdG Hamiltonian in the imaginary time, given by
\begin{eqnarray}
	Z=\int \mathcal{D}c\mathcal{D}\bar{c} e^{-S},
\end{eqnarray}
where the action is given by
\begin{eqnarray}
	S=\int_0^\beta d\tau\left( \sum_{ {\bf p},\alpha}\bar{c}_{ {\bf p}\alpha}(\partial_{\tau}-\mu)
	c_{ {\bf p}\alpha}-\hat{H}(c,\bar{c})\right).
\end{eqnarray}
In the path integral formula, $c$ and $\bar{c}$ are the Grassmann field for the operators $\hat{c}$
and $\hat{c}^{\dag}$ with $\bar{c}(\beta)=-\bar{c}(0)$ and $c(\beta)=c(0)$.

Let us assume the symmetry group for the single-particle Hamiltonian $H_0$ as $\mathsf{G}$
and we can decompose the interaction term into the representation matrices (denoted as
$\gamma_{i\mu}$ of the group $\mathsf{G}$, where $i$ labels the representation and $\mu$ labels
the dimension of the representation $i$. Let's define
\begin{eqnarray}
	\bar{B}_{i\mu}({\bf q})=\sum_{ {\bf p},\alpha\beta}\bar{c}_{ {\bf \frac{q}{2}-p},\alpha}
	(\gamma_{i\mu}({\bf p}))_{\alpha\beta}\bar{c}_{ {\bf \frac{q}{2}+p},\beta},\nonumber\\
	B_{i\mu}({\bf q})=\sum_{ {\bf p},\alpha\beta}c_{ {\bf \frac{q}{2}+p},\alpha}
	(\gamma^{\dag}_{i\mu}({\bf p}))_{\alpha\beta}c_{ {\bf \frac{q}{2}-p},\beta},\\
	V_{\alpha\beta\gamma\delta}({\bf p,p',q})=-V_i({\bf q})(\gamma_{i\mu}({\bf p}))_{\alpha\beta}
	(\gamma^{\dag}_{i\mu})
\end{eqnarray}
where the minus sign for $V_{\alpha\beta\gamma\delta}({\bf p,p',q})$ keeps $V_i({\bf q})>0$
for attractive interactions. The interaction term can be written as
\begin{eqnarray}
	H_V=-\frac{1}{2}\sum_{ {\bf q},i\mu}V_i({\bf q})\bar{B}_{i,\mu}({\bf q})B_{i\mu}({\bf q}).
\end{eqnarray}
Since
\begin{eqnarray}
	exp\left( \int d\tau \frac{1}{2}\sum_{ {\bf q},i\mu}V_i({\bf q})\bar{B}_{i\mu}B_{i\mu} \right)
	=\frac{1}{\tilde{Z}_0}
	\int\mathcal{D}\Delta^*\mathcal{D}\Delta exp\left(-\frac{1}{2}\int d\tau\sum_{ {\bf q},i\mu}
	\left( \frac{1}{V_i}\Delta^*_{i\mu}\Delta_{i\mu}+\bar{B}_{i\mu}\Delta_{i\mu}+\Delta_{i\mu}^*
	B_{i\mu} \right)	\right)
\end{eqnarray}
with $\tilde{Z}_0=\int \mathcal{D}\tilde{\Delta}^*\mathcal{D}\tilde{\Delta} exp\left(
-\int d\tau\sum_{ {\bf q},i\mu}\frac{1}{V_i}\tilde{\Delta}_{i\mu}^*\tilde{\Delta}_{i\mu}\right)$,
we may apply Hubbard-Stratonovich transformation to the above action and obtain
\begin{eqnarray}
	Z=\frac{1}{\tilde{Z}_0}\int\mathcal{D}\Delta^*\mathcal{D}\Delta\mathcal{D}\bar{c}
	\mathcal{D}c exp\left( -\int d\tau\left( \sum_{ {\bf p},\alpha\beta}\bar{c}_{ {\bf p}\alpha}
	( (\partial_\tau-\mu)\delta_{\alpha\beta}+(H_0)_{\alpha\beta})c_{ {\bf p}\beta}+\frac{1}{2}
	\sum_{ {\bf q},i\mu}\left( \frac{1}{V_i}\Delta^*_{i\mu}\Delta_{i\mu}+\bar{B}_{i\mu}\Delta_{i\mu}
	+\Delta^*_{i\mu}B_{i\mu}\right)\right) \right),
\end{eqnarray}
where $\Delta_{i\mu}$ is superconducting order parameter (gap function).
This action only contain fermion bilinear terms and thus we can integrate out the electrons
$c$ and $\bar{c}$. We write the resulting path integral as
$Z=\int \mathcal{D}\Delta^*\mathcal{D}\Delta exp\left(-S_{eff}\right)$ with
the effective action $S_{eff}=\int d\tau L_{eff}$.
Here the Lagrangian is expand as a function of the order parameters $\Delta$ and $\Delta^*$
with the form
\begin{eqnarray}
	L_{eff}=L_2+L_4, \label{Eq:App_Leff}
\end{eqnarray}
 where $L_2$ is for the second order term and $L_4$ is for the fourth order term.
 The second order term $L_2$ is given by
\begin{eqnarray}
	L_2=\frac{1}{2}\sum_{ {\bf q},i\mu}\frac{1}{V_i}\Delta^*_{i\mu}({\bf q})\Delta_{i\mu}({\bf q})-
	\frac{1}{2}\sum_{ {\bf q},i\mu,j\nu}\Delta^*_{i\mu}({\bf q})\chi^{(2)}_{ij,\mu\nu}({\bf q})\Delta_{j\nu},
	\label{Eq:App_L2}
\end{eqnarray}
where superconductivity susceptibility $\chi^{(2)}$ is given by
\begin{eqnarray}
	\chi^{(2)}_{ij,\mu\nu}({\bf q})=-\frac{1}{\beta}\sum_{i\omega_n,{\bf p}}Tr\left(\gamma^{\dag}_{i\mu}
	\mathcal{G}_e({\bf p+\frac{q}{2}},i\omega_n)\gamma_{j\nu}\mathcal{G}_h({\bf p-\frac{q}{2}},
	i\omega_n)\right).\label{Eq:App_chi2}
\end{eqnarray}
Here the single-particle Green function is defined as
$\mathcal{G}_e({\bf p},i\omega_n)=\left( i\omega_n-H_0({\bf p}) \right)^{-1}$ for electrons
and $\mathcal{G}_h({\bf p},i\omega_n)=\left( i\omega_n+H_0^*({\bf -p}) \right)^{-1}$ for holes for
our interacting Hamiltonian in the Matsubara frequency space. The fourth order term $L_4$ is given by
\begin{eqnarray}
	L_4=\frac{1}{4}\sum\Delta^*_{i\mu}({\bf q_1})\Delta^*_{i'\mu'}({\bf q_2})\chi^{(4)}_{ii',jj',\mu\mu',
	\nu\nu'}\Delta_{j\nu}({\bf q_3})\Delta_{j'\nu'}({\bf q_4})\delta_{{\bf q_1+q_2=q_3+q_4}},
	\label{Eq:App_L4}
\end{eqnarray}
where
\begin{eqnarray}
	\chi^{(4)}_{ii',jj',\mu\mu',\nu\nu'}=\frac{1}{\beta}\sum_{{\bf p},i\omega_n}Tr\left(
	\gamma^{\dag}_{i\mu}\mathcal{G}_e({\bf p},i\omega_n)\gamma_{j,\nu}\mathcal{G}_{h}({\bf p},i\omega_n)
	\gamma^{\dag}_{i',\mu'}\mathcal{G}_e({\bf p},i\omega_n)\gamma_{j',\nu'}\mathcal{G}_{h}
	({\bf p},i\omega_n)\right). \label{Eq:App_chi4}
\end{eqnarray}
Here we have neglected the spatial dependence (no ${\bf q}$ dependence)
in the fourth order superconductivity susceptibility $\chi^{(4)}$ and
only focus on the uniform case for this term.

The superconducting order parameter is determined by minimizing Ginzburg-Landau free energy, which
can be determined by considering $\frac{\delta L}{\delta \Delta^*_{i\mu}}=0$. Near the critical
temperature, the order parameter can be regarded a perturbation and in this case, we only need to consider the second order term and obtain the linearized gap equation
\begin{eqnarray}
	\Delta_{i\mu}({\bf q})=V_i\sum_{j\nu}\chi_{ij,\mu\nu}({\bf q})\Delta_{j\nu}({\bf q}).
\end{eqnarray}
for the pairing in the representation $i$. The linearized gap equation will be used to study the critical temperature of
our system.

\section{Green function and superconductivity susceptibility}
To solve the linearized gap equation, it is essential to calculate Matsubara Green functions and superconductivity
susceptibility. In this part, we consider the Hamiltonian of bilayer TMD (Eq. (1) in the main text) with Zeeman coupling,
given by
\begin{eqnarray}
	H=\xi_{\bf k}+\epsilon\beta_{so}s_z\tau_z+t\tau_z+g{\bf B}\cdot{\bf s}.
\end{eqnarray}
where $\xi_k=\frac{k^2}{2m}-\mu$ term describes kinetic energy, $\beta_{so}$ term describes Ising spin-orbit coupling
(SOC), $t$ term is for the hybridization between two layers and the last term is for Zeeman coupling with
the magnetic field ${\bf B}$.
The eigen-energy of the above Hamiltonian is given by
\begin{eqnarray}
	\varepsilon_{s\lambda}=\xi_{\bf k}+\lambda\sqrt{D_0^2+g^2B^2+2sD_1}
\end{eqnarray}
where $D_0=\sqrt{\beta^2_{so}+t^2}$ and $D_1=\sqrt{\beta^2_{so}B_z^2+t^2g^2B^2}$.
The corresponding Matsubara Green function for electrons can be written in a compact form as
\begin{eqnarray}
	\mathcal{G}_e({\bf k},i\omega_n)=\sum_{\lambda s}\frac{P^e_{\epsilon s\lambda}}
	{i\omega_n-\xi_{\bf k}-\lambda F_s}
\end{eqnarray}
where $F_s=\sqrt{D_0^2+g^2B^2+2sD_1}$ and
\begin{eqnarray}
	P^e_{\epsilon s\lambda}=\frac{1}{4}\left( 1+\frac{\lambda}{F_s}(\epsilon \beta_{so}s_z\tau_z+t\tau_x
	+g{\bf B}\cdot{\bf s}) \right)\left( 1+\frac{s}{D_1}(\epsilon \beta_{so}h_z\tau_z+tg({\bf B\cdot s})\tau_x)
	\right). 
\end{eqnarray}
The hole Green function is
\begin{eqnarray}
	\mathcal{G}_h({\bf k},i\omega_n)=\sum_{\lambda s}\frac{P^h_{\epsilon s\lambda}}{i\omega_n+\xi_{\bf k}
	-\lambda F_s}
\end{eqnarray}
where 
\begin{eqnarray}
	P^h_{\epsilon s\lambda}=\frac{1}{4}\left( 1+\frac{\lambda}{F_s}(\epsilon \beta_{so}s_z\tau_z-t\tau_x
	-g{\bf B\cdot s^T}) \right)\left( 1+\frac{s}{D_1}(-\epsilon \beta_{so}h_z\tau_z+tg({\bf B\cdot s^T})\tau_x)
	\right).
\end{eqnarray}

Now we only focus on the uniform case (${\bf q}=0$) and may substitute the form of electron and hole Green functions into superconductivity susceptibility to obtain
\begin{eqnarray}
	\chi^{(2)}_{ij,\mu\nu}(0)=-\frac{1}{\beta}
	\sum_{\epsilon,\lambda\lambda'ss'}\sum_{i\omega_n,{\bf k}}
	\frac{1}{i\omega_n-\xi_{\bf k}-\lambda F_s}\frac{1}{i\omega_n+\xi_{\bf k}-\lambda'F_{s'}}
	Tr\left(\gamma^{\dag}_{i\mu}P^e_{\epsilon s\lambda}\gamma_{j\nu}
	P^h_{\epsilon s'\lambda'}\right).
\end{eqnarray}
Since $Tr\left(\gamma^{\dag}_{i\mu}P^e_{\epsilon s\lambda}\gamma_{j\nu}P^h_{\epsilon s'\lambda'}\right)$ does not depend on $i\omega_n$ and ${\bf k}$, we can fisrt integrate out $\sum_{i\omega_n,{\bf k}}\frac{1}{i\omega_n-\xi_{\bf k}-\lambda F_s}\frac{1}{i\omega_n+\xi_{\bf k}-\lambda'F_{s'}}$. This integral can be further simplified. One can show that the singular behavior of the above expression comes from the term $\lambda=-\lambda'$ when $D_0$ is large. Therefore, we consider the case with $\lambda=-\lambda'=-$ (the case with $\lambda=-\lambda'=+$ is similar). In this case, we define
\begin{eqnarray}
	\chi_{0,ss'}=-\frac{1}{\beta}\sum_{i\omega_n,{\bf k}}\frac{1}{i\omega_n-\xi_{\bf k}+F_s}\frac{1}{i\omega_n+\xi_{\bf k}-F_{s'}}
\end{eqnarray}
and direct calculation shows that
\begin{eqnarray}
	\chi_{0,ss'}=N_0 ln\left( \frac{2\gamma\omega_D}{\pi kT} \right)
	+N_0 \psi\left( \frac{1}{2} \right)-\frac{N_0}{2}\left( \psi\left( \frac{1}{2}
	-\frac{i}{2\pi kT}(F_s-F_{s'})\right) +\psi \left( \frac{1}{2}+\frac{i}{2\pi kT}
	(F_s-F_{s'})\right)\right)
\end{eqnarray}
where $N_0$ is density of states at the Fermi energy, $\gamma=1.57$,
$\omega_D$ is Debye frequency, which is chosen to be $10meV$ 
in our calculation, and $\psi$ is the di-gamma function. Therefore, we have
$\chi_{0,++}=\chi_{0,--}=N_0 ln\left( \frac{2\gamma\omega_D}{\pi kT} \right)=\chi_0$
and $\chi_{0,+-}=\chi_{0,-+}=\chi_0-\delta\chi$ with
\begin{eqnarray}
	&&\delta\chi=\frac{N_0}{2}\left( \psi\left( \frac{1}{2}
	-\frac{i}{2\pi kT}(F_s-F_{s'})\right) +\psi \left( \frac{1}{2}+\frac{i}{2\pi kT}
	(F_s-F_{s'})\right)-2\psi\left( \frac{1}{2} \right) \right)\approx
	\frac{C}{2}\left( \frac{1}{2\pi kT} \right)^2\left( F_+-F_{-} \right)^2\nonumber\\
	&&\approx 2C\left( \frac{1}{2\pi kT} \right)^2\frac{D_1^2}{D_0^2}\propto B^2,
\end{eqnarray}
where $C=-\psi^{(2)}\left( \frac{1}{2} \right)\approx 16.83$.
With the above expression, the superconductivity susceptibility is simplified as
\begin{eqnarray}
	\chi^{(2)}_{ij,\mu\nu}=\sum_{\epsilon,ss'}\chi_{0,ss'}
	Tr\left(\gamma^{\dag}_{i\mu}P^e_{\epsilon s-}\gamma_{j\nu}
	P^h_{\epsilon s'+}\right)
\end{eqnarray}
for $\lambda=-\lambda'=-$. 

To obtain the gap equation, we need further to decompose the pairing interaction
into different representations. 
The pairing interaction is introduced as $\hat{H}_V=\frac{1}{2}\sum_{p,p',q}V_{\sigma_1\sigma_2\sigma_3\sigma_4}^{\eta_1\eta_2\eta_3\eta_4}
c^{\dag}_{\sigma_1,\eta_1,\frac{q}{2}-p}c^{\dag}_{\sigma_2,\eta_2,\frac{q}{2}+p}c_{\sigma_3,\eta_3,\frac{q}{2}+p'}
c_{\sigma_4,\eta_4,\frac{q}{2}-p'}$. For simplicity, we assume the interaction is momentum independent (on-site interaction)
 and only consider the following non-zero $V_{\sigma_1\sigma_2\sigma_3\sigma_4}^{\eta_1\eta_2\eta_3\eta_4}$:
 (1) intra-layer interaction 
 \begin{eqnarray}
	V_{\sigma\bar{\sigma}\bar{\sigma}\sigma}^{\eta\eta\eta\eta}=
	-V_{\sigma\bar{\sigma}\sigma\bar{\sigma}}^{\eta\eta\eta\eta}=-U_0 
	\label{App_Eq:U0}
 \end{eqnarray}
 and (2) inter-layer interaction 
\begin{eqnarray}
V_{\sigma\sigma\sigma\sigma}^{\eta\bar{\eta}\bar{\eta}\eta}=-V_{\sigma\sigma\sigma\sigma}^{\eta\bar{\eta}\eta\bar{\eta}}
=V_{\sigma\bar{\sigma}\bar{\sigma}\sigma}^{\eta\bar{\eta}\bar{\eta}\eta}
=-V_{\sigma\bar{\sigma}\sigma\bar{\sigma}}^{\eta\bar{\eta}\eta\bar{\eta}}=-V_0,
\label{App_Eq:V0}
 \end{eqnarray}
where $\bar{\sigma}$ and $\bar{\eta}$ reverse the value of
$\sigma$ and $\eta$. Here we introduce additional minus sign before $U_0$ and $V_0$ so that the attractive interaction is defined for $U_0>0$ and $V_0>0$. 
Next we decompose the interaction into different channels 
with the form 
\begin{eqnarray}
	V_{\sigma_1\sigma_2\sigma_3\sigma_4}^{\eta_1\eta_2\eta_3\eta_4}=-\sum_{i,\mu}V_{i,\mu}(\gamma_{i,\mu})_{\sigma_1\sigma_2,\eta_1\eta_2}
	(\gamma_{i,\mu}^{\dag})_{\sigma_3\sigma_4,\eta_3\eta_4}\label{App_Eq:Vint},
\end{eqnarray}
where $\gamma$ matrices label different representation matrices. 
For the $D_{3d}$ group,
$s_y$ and $s_y\tau_x$ (labeled as $\gamma_{A_{1g},1}$ and $\gamma_{A_{1g},2}$) belong to the $A_{1g}$ representation,
$s_y\tau_z$ (labeled as $\gamma_{A_{1u}}$) belongs to $A_{1u}$,
$s_x\tau_y$ (labeled as $\gamma_{A_{2u}}$) belongs to $A_{2u}$ and $\{\tau_y,s_z\tau_y\}$
(labeled as $\{\gamma_{E_u,1},\gamma_{E_u,2}\}$) to $E_u$.
More explicitly, $\gamma_{A_{1g},1}$ and $\gamma_{A_{1g},2}$ describe intra-layer and inter-layer singlet pairings
($c_{\uparrow+}c_{\downarrow+}+c_{\uparrow-}c_{\downarrow-}$ and
$c_{\uparrow+}c_{\downarrow-}+c_{\uparrow-}c_{\downarrow+}$), $\gamma_{A_{1u}}$ corresponds to intra-layer singlet
pairings with opposite phases between two layers ($c_{\uparrow+}c_{\downarrow+}-c_{\uparrow-}c_{\downarrow-}$), $\gamma_{A_{2u}}$ is for inter-layer singlet pairing ($c_{\uparrow+}c_{\downarrow-}-c_{\uparrow-}c_{\downarrow+}$)
and $(\gamma_{E_u,1},\gamma_{E_u,2})$ gives inter-layer triplet pairing ($c_{\uparrow+}c_{\uparrow-}$
and $c_{\downarrow+}c_{\downarrow-}$).
With these matrices, we can compare matrix elements of interactions
in Eq. (\ref{App_Eq:Vint}) with those in Eq. (\ref{App_Eq:U0})
and Eq. (\ref{App_Eq:V0}) and obtain
$V_{A_{1g},1}=V_{A_{1u}}=\frac{U_0}{2}$ and
$V_{A_{1g},2}=V_{A_{2u}}=V_{E_u,1}=V_{E_u,2}=\frac{V_0}{2}$. 
 
Next we need to evaluate the element
$Tr\left(\gamma^{\dag}_{i\mu}P^e_{\epsilon s-}\gamma_{j\nu}P^h_{\epsilon s'+}\right)$
and discuss the gap equation in different representations separately.

(1) $A_{1g}$ representation ($s_y, s_y\tau_x$)

Since we have two possible representation matrices for the $A_{1g}$ representation,
the gap function can be expanded as $\Delta_{A_{1g}}=\Delta_{A_{1g},1}s_y+\Delta_{A_{1g},2}
s_y\tau_x$. We can compute $\chi^{(2)}_{A_{1g},\mu\nu}$ directly and only keep terms up to the
second order in ${\bf B}$. The superconductivity susceptibility is given by
\begin{eqnarray}
	&&\chi_{A_{1g},11}=4(\chi_0-\delta\chi)\\
	&&\chi_{A_{1g},12}=\chi_{A_{1g},21}=2(\chi_0-\delta\chi)\left( -\frac{t}{D_0}-
	\frac{t^3g^2B^2}{D_0D_1^2}\right)+2\chi_0\left( -\frac{t}{D_0}+\frac{t^3g^2B^2}{D_0D_1^2}
	\right)\\
	&&\chi_{A_{1g},22}=2\chi_0\left( \frac{t^2}{D_0^2}-\frac{t^2g^2}{D_1^2}\left(
	B_z^2+\frac{t^2}{D_0^2}B^2_{\parallel}\right) \right)+2(\chi_0-\delta\chi)\left(
	\frac{t^2}{D_0^2}+\frac{t^2g^2}{D_1^2}\left( B_z^2+\frac{t^2}{D_0^2}B^2_{\parallel}
	\right)\right),
\end{eqnarray}
where $\chi_{A_{1g},11}$ is for $s_y$, $\chi_{A_{1g},22}$ is for $s_y\tau_x$ and
$\chi_{A_{1g},12}$ and $\chi_{A_{1g},21}$ describe the coupling between $s_y$
and $s_y\tau_x$. Since $V_{A_{1g},1}=\frac{U_0}{2}$ and $V_{A_{1g},2}=\frac{V_0}{2}$,
we obtain a set of linear equations
\begin{eqnarray}
	&&\Delta_{A_{1g},1}=\frac{U_0}{2}\left( \chi^{(2)}_{A_{1g},11}\Delta_{A_{1g},1}
	+\chi^{(2)}_{A_{1g},12}\Delta_{A_{1g},2}\right),\nonumber\\
	&&\Delta_{A_{1g},2}=\frac{V_0}{2}\left( \chi^{(2)}_{A_{1g},21}\Delta_{A_{1g},1}
	+\chi^{(2)}_{A_{1g},22}\Delta_{A_{1g},2}\right). \label{Eq:App_A1g_eq}
\end{eqnarray}

Let's first discuss the case without magnetic field ${\bf B}=0$ and in this case, the gap equations
are $\frac{\Delta_{A_{1g},1}}{\chi_0}=2U_0\left( \Delta_{A_{1g},1}-\frac{t}{D_0}\Delta_{A_{1g},2}
\right)$ and $\frac{\Delta_{A_{1g},2}}{\chi_0}=2V_0\left( -\frac{t}{D_0}\Delta_{A_{1g},1}
+\frac{t^2}{D_0^2}\Delta_{A_{1g},2}\right)$. The above equations can be viewed as an eigen-equation
for $\Delta_{A_{1g},1}$ and $\Delta_{A_{1g},2}$ and its eigen-values are $\frac{1}{\chi_0}=0$
and $\frac{1}{\chi_0}=U_0+V_0\frac{t^2}{D_0^2}$. With the expression of $\chi_0$, we find the first
eigen-solution corresponds to vanishing $T_c$ and the second eigen-solution gives rise to the
critical temperature
\begin{equation}
	kT_{c0,A_{1g}}=\frac{2\gamma\omega_D}{\pi} exp\left( -\frac{1}{2N_0 (U_0+V_0\frac{t^2}{D_0^2})} \right).
\end{equation}
The corresponding eigen-state of $A_{1g}$ pairing satisfies
\begin{eqnarray}
	\Delta_{A_{1g},2}=-\frac{V_0t}{U_0D_0}\Delta_{A_{1g},1}. \label{Eq:App_A1g}
\end{eqnarray}
Therefore in case $t\ll D_0$, the $\Delta_{A_{1g},1}$ part dominates.

For a non-zero but weak ${\bf B}$, we have $\delta\chi\ll\chi_0$ and the critical temperature is close to $T_{c0,A_{1g}}$.
The correction to the critical temperature can be obtained by substituting (\ref{Eq:App_A1g}) into (\ref{Eq:App_A1g_eq})
and is determined by
\begin{eqnarray}
	N_0 ln\left( \frac{T_{c,A_{1g}}}{T_{c0,A_{1g}}} \right)=-\frac{2U_0+V_0\left( \frac{t^2}{D_0^2}
	+\frac{t^4B^2}{D_0^2D_1^2} \right)}{2U_0+2V_0\frac{t^2}{D_0^2}}\delta\chi.
\end{eqnarray}
For an in-plane magnetic field ($D_1=tB_{\parallel}$), we have
\begin{eqnarray}
	ln\left( \frac{T_{c\parallel,A_{1g}}}{T_{c0,A_{1g}}} \right)\approx
	-2C\left(
	\frac{1}{2\pi kT_{c0,A_{1g}}} \right)^2\frac{t^2}{\beta^2_{so}}g^2B^2_{\parallel},
\end{eqnarray}
while for an out-of-plane magnetic field ($D_1=\beta_{so}B_z$), we have
\begin{eqnarray}
	ln\left( \frac{T_{cz,A_{1g}}}{T_{c0,A_{1g}}} \right)\approx-2C\left(
	\frac{1}{2\pi kT_{c0,A_{1g}}} \right)^2g^2B_z^2.
\end{eqnarray}
Comparing the critical temperatures for in-plane magnetic fields and out-of-plane magnetic fields,
we find that the paramagnetic effect for in-plane magnetic fields is much weaker than that
for out-of-plane magnetic fields due to the factor $\frac{t^2}{\beta^2_{so}}$, which will lead to
the high in-plane critical magnetic field and is consistent
with the experimental observations in bilayer TMD materials.

(2) $A_{1u}$ pairing ($s_y\tau_z$)

The superconductivity susceptibility is given by
\begin{eqnarray}
	\chi^{(2)}_{A_{1u}}=4\chi_0\frac{\beta_{so}^2}{D_0^2}-\delta\chi\left(
	\frac{2\beta_{so}^2}{D_0^2}+\frac{\beta^2_{so}g^2B_z^2-t^2g^2B^2}{D_1^2}+\frac{
	D_0^4g^2B_z^2-\beta_{so}^2t^2g^2B_{\parallel}^2+t^4g^2B_{\parallel}^2}{D_0^2D_1^2}\right)
\end{eqnarray}
and the corresponding gap equation is $\Delta_{A_{1u}}=\frac{U_0}{2}\chi^{(2)}_{A_{1u}}\Delta_{A_{1u}}$.
At zero magnetic field ${\bf B}=0$ and $\delta\chi=0$, we find
\begin{eqnarray}
	kT_{c0,A_{1u}}=\frac{2\gamma\omega_D}{\pi} exp\left( -\frac{D_0^2}{2U_0\beta^2_{so}} \right).
\end{eqnarray}

In a weak magnetic field and in the limit $t\ll\beta_{so}$,
\begin{eqnarray}
	ln\left( \frac{T_{c,A_{1u}}}{T_{c0,A_{1u}}} \right)=-\left( 2+\frac{1}{D_1^2}\left(
	2\beta_{so}^2g^2B_z^2-2t^2g^2B^2_{\parallel}-t^2g^2B_z^2+\frac{t^4}{\beta_{so}^2}g^2B_{\parallel}^2
	\right) \right)\frac{\delta\chi}{4}.
\end{eqnarray}
Thus, for out-of-plane magnetic fields,
\begin{eqnarray}
	ln\left( \frac{T_{cz,A_{1u}}}{T_{c0,A_{1u}}} \right)=-2C\left( \frac{1}{2\pi kT_{c0,A_{1u}}}
	\right)^2g^2B_z^2,
\end{eqnarray}
and for in-plane magnetic fields,
\begin{eqnarray}
	N_0 ln\left( \frac{T_{c\parallel,A_{1u}}}{T_{c0,A_{1u}}} \right)=-\frac{C}{2}
	\left( \frac{1}{2\pi kT_{c0,A_{1u}}}\right)^2\frac{t^4}{\beta_{so}^4}g^2B_{\parallel}^2.
\end{eqnarray}
We find a factor $\frac{t^4}{\beta_{so}^4}$ for in-plane magnetic fields but not for out-of-plane magnetic fields.

(3) $A_{2u}$ pairing ($s_x\tau_y$)

In this case, we find that $\chi^{(2)}_{A_{2u}}=0$ and thus no superconducting phase is possible for this
representation.

(4) $E_{u}$ pairing ($\left\{ \tau_y, s_z\tau_y \right\}$)

Since the $E_u$ pairing are two dimensional, we can write down a gap equation for each component. However, since they
are related to each other by symmetry, we expect two components share the same $T_c$. Therefore, we only consider
$\tau_y$ part here. The superconductivity susceptibility is
\begin{eqnarray}
	\chi^{(2)}_{E_u}=4\chi_0\frac{\beta^2_{so}}{D_0^2}-\left(
	2\frac{\beta^2_{so}}{D_0^2}-\frac{1}{D_1^2}\left( (\beta^2_{so}+D_0^2)g^2B_z^2-2t^2g^2B^2_{\parallel}
	-t^2g^2B_z^2 \right)\right)\delta\chi.
\end{eqnarray}
At zero magnetic field, we have
\begin{eqnarray}
	kT_{c0,E_u}=\frac{2\gamma\omega_D}{\pi} exp\left( -\frac{D_0^2}
	{2N_0V_0\beta^2_{so}} \right).
\end{eqnarray}

For a weak magnetic field and $t\ll\beta_{so}$, we have
\begin{eqnarray}
	N_0 ln\left( \frac{T_{c,E_{u}}}{T_{c0,E_{u}}} \right)=-\left( 2-\frac{D_0^2g^2}{D_1^2}
	\left( B_z^2-\frac{t^2}{\beta^2_{so}}B^2 \right)-\frac{D_0^2g^2}{D_1^2\beta^2_{so}}
	\left( D_0^2B_z^2-t^2B_{\parallel}^2 \right)\right)\frac{\delta\chi}{4}.
\end{eqnarray}

For out-of-plane magnetic fields,
\begin{eqnarray}
	ln\left( \frac{T_{cz,E_u}}{T_{c0,E_u}} \right)=0,
\end{eqnarray}
and for in-plane magnetic fields,
\begin{eqnarray}
	ln\left( \frac{T_{c\parallel,E_u}}{T_{c0,E_u}} \right)=-C
	\left( \frac{1}{2\pi kT_{c0,A_{1u}}}\right)^2
	\left( 1+\frac{\beta^2_{so}}{D_0^2} \right)\frac{t^2}{\beta^2_{so}}g^2B^2_{\parallel}.
\end{eqnarray}
Therefore, we find Zeeman coupling will not reduce the $T_c$ for out-of-plane magnetic fields
and the contribution of in-plane magnetic fields has a factor of
$\left( 1+\frac{\beta^2_{so}}{D_0^2} \right)\frac{t^2}{\beta^2_{so}}$.
This is because the $E_u$ pairing corresponds to the interlayer equal spin pairing.

\section{Ginzburg-Landau free energy}
In the above, we have presented our derivations and results of the linearized gap equation
for bilayer TMD materials. In this section, we will construct the Ginzburg-Landau free energy
for our system. In particular, this approach will allow us to study inhomogeneous superconductivity
when we consider the orbital effect of magnetic fields.

The Ginzburg-Landau free energy is given by Eqs. (\ref{Eq:App_Leff})-(\ref{Eq:App_chi4}), in which one needs to
evaluate superconductivity susceptibility $\chi^{(2)}$ and $\chi^{(4)}$. We have computed $\chi^{(2)}$ for ${\bf q}=0$,
which can be directly applied to Landau free energy, in the last section for the linearized gap equation.
In this part, we need to further include the ${\bf q}$ dependence in order to discuss the gradient term
in the Landau free energy.

With a finite ${\bf q}$, the superconductivity susceptibility $\chi^{(2)}$ is changed to
\begin{eqnarray}
	\chi^{(2)}_{ij,\mu\nu}({\bf q})=-\frac{1}{\beta}\sum_{ss',\lambda\lambda'}\sum_{i\omega_n,\epsilon,{\bf k}}
	\frac{1}{i\omega_n-\xi_{\bf k+\frac{q}{2}}-\lambda F_s}\frac{1}{i\omega_n+\xi_{\bf k-\frac{q}{2}}-\lambda'F_{s'}}
	Tr\left(\gamma^{\dag}_{i\mu}P^e_{\epsilon s\lambda}\gamma_{j\nu}
	P^h_{\epsilon s'\lambda'}\right).
\end{eqnarray}
There is no momentum dependence in $Tr\left(\gamma^{\dag}_{i\mu}P^e_{\epsilon s\lambda}\gamma_{j\nu}
P^h_{\epsilon s'\lambda'}\right)$ and thus, we only need to consider $\xi_{\bf k\pm\frac{q}{2}}$. We may treat
${\bf q}$ as a small number and expand it as $\xi_{\bf k\pm\frac{q}{2}}=\xi_{\bf k}\pm \frac{\hbar }{2}{\bf v_k}\cdot
{\bf q}$ up to the linear term in ${\bf q}$. Since we have already discussed the Zeeman coupling, we will neglect
this term in the discussion below for simplicity ($F_s=D_0$). In this case, we define
\begin{eqnarray}
	\chi_{0,\lambda\lambda'}({\bf q})=-\frac{1}{\beta}\sum_{i\omega_n,{\bf k}}
	\frac{1}{i\omega_n-\xi_{\bf k+\frac{q}{2}}-\lambda D_0}
	\frac{1}{i\omega_n+\xi_{\bf k-\frac{q}{2}}-\lambda'D_0},
\end{eqnarray}
which is independent of $s$ and $s'$. $\chi^{(2)}_{ij,\mu\nu}({\bf q})=\sum_{ss',\lambda\lambda'}
\chi_{0,\lambda\lambda'}({\bf q})Tr\left(\gamma^{\dag}_{i\mu}P^e_{\epsilon s\lambda}\gamma_{j\nu}
P^h_{\epsilon s'\lambda'}\right)$. Furthermore, we only focus on $\lambda=-\lambda'=-$ and in this case,
\begin{eqnarray}
	\chi_0({\bf q})=\chi_{0,-+}({\bf q})=
	N_0 ln\left( \frac{2\gamma\omega_D}{\pi kT} \right)+N_0\psi\left( \frac{1}{2} \right)
	-\frac{N_0}{8\pi}\int d\Omega_k \left( \psi \left( \frac{1}{2}-\frac{i}{2\pi kT}\hbar v_f {\bf e_k\cdot q}
	\right)+\psi\left( \frac{1}{2}+\frac{i}{2\pi kT}\hbar v_f {\bf e_k\cdot q} \right) \right),
\end{eqnarray}
where $v_f$ is the Fermi velocity, ${\bf e_k}$ is the unit vector of the momentum ${\bf k}$ and
$\Omega_k$ is the solid angle for the ${\bf k}$ integral. By expand the digamma function and perform the
$\Omega_k$ integral, we obtain
\begin{eqnarray}
	\chi_0({\bf q})=N_0 ln\left( \frac{2\gamma\omega_D}{\pi kT} \right)-\frac{N_0C}{6}\left(
	\frac{\hbar v_f}{2\pi kT} \right)^2q^2.
\end{eqnarray}
From the above equation, we can construct the second order term $L_2$ in the Landau free energy. We will next
discuss the pairing in each representation, separately. The $T$ in the above expression should be replaced by
the $T_{c0}$ for the corresponding pairing.

For the fourth order term, we need to evaluate $\chi^{(4)}$, which is written as
\begin{eqnarray}
	&&\chi^{(4)}_{ii',jj',\mu\mu',\nu\nu'}=\frac{1}{\beta}\sum_{\epsilon, \{s,\lambda\}}
	\sum_{{\bf k},i\omega_n}\frac{1}{i\omega_n-\xi_{\bf k}-\lambda_1D_0}\frac{1}{i\omega_n
	+\xi_{\bf k}-\lambda_2D_0}\frac{1}{i\omega_n-\xi_{\bf k}-\lambda_3D_0}\frac{1}{i\omega_n
	+\xi_{\bf k}-\lambda_4D_0}\nonumber\\
	&&Tr\left(
	\gamma^{\dag}_{i\mu}P^e_{\epsilon s_1\lambda_1}\gamma_{j,\nu}P^h_{\epsilon s_2\lambda_2}
	\gamma^{\dag}_{i',\mu'}P^e_{\epsilon s_3\lambda_3}\gamma_{j',\nu'}P^h_{\epsilon s_4\lambda_4}
	\right).
\end{eqnarray}
We again only consider the most singular part of the momentum-frequency space integral, which is contributed
from $\lambda_1=\lambda_3=-$ and $\lambda_2=\lambda_4=+$. Therefore, we have
\begin{eqnarray}
	&&\chi^{(4)}_{ii',jj',\mu\mu',\nu\nu'}=\frac{7N_0\zeta(3)}{8(\pi kT)^2}\sum_{\epsilon, \{s\}}
	Tr\left( \gamma^{\dag}_{i\mu}P^e_{\epsilon s_1-}\gamma_{j,\nu}P^h_{\epsilon s_2+}
	\gamma^{\dag}_{i',\mu'}P^e_{\epsilon s_3-}\gamma_{j',\nu'}P^h_{\epsilon s_4+}
	\right),\label{eq:APP_fourth}
\end{eqnarray}
where $\zeta(x)$ is the zeta-function.

(1) $A_{1g}$ pairing

Direct calculation of superconductivity susceptibility almost recovers our previous results with the replacement
of $\chi_0$ by $\chi_0({\bf q})$. Therefore, we have $\chi_{A_{1g},11}=4\chi_0({\bf q})$, $\chi_{A_{1g},12}=
\chi_{A_{1g},21}=-\frac{4t}{D_0}\chi_0({\bf q})$ and $\chi_{A_{1g},22}=\frac{4t^2}{D_0^2}\chi_0({\bf q})$. Therefore,
the second order term of Landau free energy is written as
\begin{eqnarray}
	&&L_{2,A_{1g}}=\frac{1}{2}\sum_{\bf q} \left(
	\begin{array}{cc}
		\Delta^*_{A_{1g},1}&\Delta^*_{A_{1g},2}
	\end{array}
	\right)
	\left(
	\begin{array}{cc}
		\frac{1}{V_{A_{1g},1}}-\chi_{A_{1g},11}&-\chi_{A_{1g},12}\\
		-\chi_{A_{1g},21}&\frac{1}{V_{A_{1g},2}}-\chi_{A_{1g},22}
	\end{array}
	\right)
	\left(
	\begin{array}{c}
		\Delta_{A_{1g},1}\\
		\Delta_{A_{1g},2}
	\end{array}
	\right)\nonumber\\
        &&=\sum_{\bf q} \left(
	\begin{array}{cc}
		\Delta^*_{A_{1g},1}&\Delta^*_{A_{1g},2}
	\end{array}
	\right)
	\left(
	\begin{array}{cc}
		\frac{1}{U_0}-2\chi_0({\bf q})&\frac{2t}{D_0}\chi_0({\bf q})\\
		\frac{2t}{D_0}\chi_0({\bf q})&\frac{1}{V_0}-2\chi_0({\bf q})
	\end{array}
	\right)
	\left(
	\begin{array}{c}
		\Delta_{A_{1g},1}\\
		\Delta_{A_{1g},2}
	\end{array}
	\right)
\end{eqnarray}
One can take the derivative of $L_{2,A_{1g}}$ with respect to $\Delta^*_{A_{1g},i}$ and
recover the linearized gap equation. Since the $\Delta_{A_{1g},1}$ pairing dominates when
$t\ll\beta_{so}$, we may substitute $\Delta_{A_{1g},2}$ with $\Delta_{A_{1g},1}$ and obtain
\begin{eqnarray}
	L_{2,A_{1g}}=\int d{\bf r} \left( \mathcal{K}_{A_{1g}}\left|\nabla \Delta_{A_{1g},1}({\bf r})\right|^2
	+\mathcal{A}_{A_{1g}}\left|\Delta_{A_{1g},1}({\bf r})\right|^2\right)\label{Eq:App_L2A1g}
\end{eqnarray}
with $\mathcal{K}_{A_{1g}}=\frac{N_0C}{3}\left( \frac{\hbar v_f}{2\pi kT} \right)^2\left( 1
+\frac{V_0t^2}{U_0D_0^2}\right)^2$ and $\mathcal{A}_{A_{1g}}=2N_0\left( 1
+\frac{V_0t^2}{U_0D_0^2}\right)^2 ln\left( \frac{T}{T_{c0,A_{1g}}} \right)\approx
2N_0\left( 1+\frac{V_0t^2}{U_0D_0^2}\right)^2 \frac{T-T_{c0,A_{1g}}}{T_{c0,A_{1g}}}$ when
the temperature is close to $T_{c0,A_{1g}}$. We further label $\mathcal{C}_{A_{1g}}=2N_0\left( 1
+\frac{V_0t^2}{U_0D_0^2}\right)^2$ and $\mathcal{A}_{A_{1g}}=\mathcal{C}_{A_{1g}}
ln\left( \frac{T}{T_{c0,A_{1g}}} \right)$.

The fourth order term can also be computed directly with the electron and hole Green functions, and
we obtain
\begin{eqnarray}
	L_{4,A_{1g}}=\int d{\bf r}\left( \mathcal{B}_{A_{1g}}|\Delta_{A_{1g,1}}({\bf r})|^4 \right),
	\label{Eq:App_L4A1g}
\end{eqnarray}
where $\mathcal{B}_{A_{1g}}=\frac{7N_0\zeta(3)}{8(\pi kT)^2}$.
Since we only concerns the temperature close to the critical temperature, $T$ in $\mathcal{K}$ and
$\mathcal{B}$ can be replaced by $T_{c0,A_{1g}}$. Eqs. (\ref{Eq:App_L2A1g}) and (\ref{Eq:App_L4A1g})
together form the Landau free energy for the $A_{1g}$ pairing.

For the out-of-plane magnetic field, the orbital effect can be taken into account by replacing
$-i\nabla \rightarrow -i\nabla+\frac{2e}{\hbar}{\bf A}$, where ${\bf A}$ is the gauge potential.
The orbital effect of in-plane magnetic fields will be discussed in details later.

(2) $A_{1u}$ pairing

The second order term in the Landau free energy for the $A_{1u}$ pairing is given by
\begin{eqnarray}
	L_{2,A_{1u}}=\int d{\bf r} \left( \mathcal{K}_{A_{1u}}\left|\nabla \Delta_{A_{1u}}({\bf r})\right|^2
	+\mathcal{A}_{A_{1u}}\left|\Delta_{A_{1u}}({\bf r})\right|^2\right)\label{Eq:App_L2A1u},
\end{eqnarray}
where $\mathcal{K}_{A_{1u}}=\frac{N_0C}{3}\left( \frac{\hbar v_f}{2\pi kT} \right)^2\frac{\beta_{so}^2}{D_0^2}$,
$\mathcal{A}_{A_{1u}}=\frac{2N_0\beta_{so}^2}{D_0^2} ln\left( \frac{T}{T_{c0,A_{1u}}} \right)\approx
\frac{2N_0\beta_{so}^2}{D_0^2}\frac{T-T_{c0,A_{1u}}}{T_{c0,A_{1u}}}$ and $\mathcal{C}=\frac{2N_0\beta_{so}^2}{D_0^2}$.

The fourth order term is given by
\begin{eqnarray}
	L_{4,A_{1u}}=\int d{\bf r}\left( \mathcal{B}_{A_{1u}}|\Delta_{A_{1u}}({\bf r})|^4 \right),
	\label{Eq:App_L4A1u}
\end{eqnarray}
where $\mathcal{B}_{A_{1u}}=\frac{7N_0\zeta(3)\beta_{so}^4}{8(\pi kT)^2D_0^4}$.

(3) $E_u$ pairing

The second order term and the fourth order term are given by
\begin{eqnarray}
	L_{2,E_{u}}=\int d{\bf r} \sum_{\mu}\left( \mathcal{K}_{E_{u}}\left|\nabla \Delta_{E_{u},\mu}({\bf r})\right|^2
	+\mathcal{A}_{E_{u}}\left|\Delta_{E_{u},\mu}({\bf r})\right|^2\right)\label{Eq:App_L2Eu},
\end{eqnarray}
and
\begin{eqnarray}
	L_{4,E_{u}}=\int d{\bf r}\left( \mathcal{B}_{E_{u},1} (|\Delta_{E_u,1}|^2+|\Delta_{E_u,2}|^2)^2
	+\mathcal{B}_{E_u,2} (\Delta_{E_u,1}^*\Delta_{E_u,2}-\Delta_{E_u,1}\Delta_{E_u,2}^*)^2\right),
	\label{Eq:App_L4Eu}
\end{eqnarray}
where $\mathcal{K}_{E_u}=\frac{N_0C}{3}\left( \frac{\hbar v_f}{2\pi kT} \right)^2\frac{\beta_{so}^2}{D_0^2}$,
$\mathcal{A}_{E_u}=\frac{2N_0\beta_{so}^2}{D_0^2} ln\left( \frac{T}{T_{c0,E_u}} \right)\approx
\frac{2N_0\beta_{so}^2}{D_0^2}\frac{T-T_{c0,E_u}}{T_{c0,E_u}}$, $\mathcal{C}_{E_u}=\frac{2N_0\beta_{so}^2}{D_0^2}$,
$\mathcal{B}_{E_{u},1}=\frac{7N_0\zeta(3)\beta_{so}^4}{8(\pi kT)^2D_0^4}$
and $\mathcal{B}_{E_{u},2}=-\frac{7N_0\zeta(3)\beta_{so}^4}{8(\pi kT)^2D_0^4}$ in the weak coupling limit.
It is known that when $\mathcal{B}_{E_{u},2}<0$, the nematic superconductivity with pairing function
$(\Delta_{E_u,1},\Delta_{E_u,2})=\Delta_{E_u}(\cos\theta,\sin\theta)$ will become stable.
On the other hand, if $\mathcal{B}_{E_{u},2}>0$, chiral superconductivity
$(\Delta_{E_u,1},\Delta_{E_u,2})=\Delta_{E_u}(1,i)$ will be realized.

\section{The orbital effect of in-plane magnetic fields}
In this part, we will discuss the orbital effect of in-plane magnetic fields, 
which turns out to be important
for inducing inhomogeneous superconducting pairing. 
Normally, the orbital effect of in-plane magnetic fields
is neglected in 2D systems because of quantum confinement 
along the out-of-plane direction. However, as we will
show below, it will have an interesting consequence in 2D TMD materials 
due to the unique band structures.

Let us assume the magnetic field is along the $x$ direction and the corresponding gauge potential is chosen as
${\bf A}=(0,-B_xz,0)$. The distance between two layers of TMD materials is taken as $z_0$ and the origin point
is chosen at the center between two layers. Thus, in our Hamiltonian, we need to replace $\xi_{\bf k}$ by
$\xi_{\bf \pi}=\frac{\hbar^2}{2m}(k_x^2+(k_y-Q\tau_z)^2)-\mu$, where $Q=\frac{eB_xz_0}{2\hbar}$. We may expand
$\xi_{\bf \pi}$ and keep only the first order term in $B_x$. The resulting Hamiltonian is
\begin{eqnarray}
	H=\xi_{\bf k}-\hbar v_Qk_y\tau_z+\epsilon \beta_{so}s_z\tau_z+t\tau_x,
\end{eqnarray}
with $v_Q=\frac{eB_x z_0}{2m}$ and the corresponding energy dispersion is given by
\begin{eqnarray}
	\varepsilon_{s\lambda}(\epsilon,{\bf k})=\xi_{\bf k}+\lambda F_{\epsilon s}(k_y),
\end{eqnarray}
where $F_{\epsilon s}=\sqrt{(\beta_{so}-\epsilon s\hbar v_Q k_y)^2+t^2}$. The energy dispersion is shown in Fig. 1d in the main text for the limit $t=0$.

The electron and hole Green functions are given by
\begin{eqnarray}
	&&\mathcal{G}_e(\epsilon,{\bf k},i\omega_n)=\sum_{\lambda s}\frac{P^e_{\epsilon s\lambda}(k_y)}
	{i\omega_n-\xi_{\bf k}-\lambda F_{\epsilon s}(k_y)}\\
	&&\mathcal{G}_h(\epsilon,{\bf k},i\omega_n)=\sum_{\lambda s}\frac{P^h_{\epsilon s\lambda}(k_y)}
	{i\omega_n+\xi_{\bf k}-\lambda F_{\epsilon s}(k_y)}
\end{eqnarray}
where we have used $\xi_{\bf -k}=\xi_{\bf k}$ and
\begin{eqnarray}
	&&P^e_{\epsilon s\lambda}=\frac{1}{4}\left( 1+\frac{\lambda}{F_{\epsilon s}(k_y)}(-\hbar v_Qk_y \tau_z
	+\epsilon \beta_{so}s_z\tau_z+t\tau_x) \right)\left( 1+s s_z \right)\\
	&&P^h_{\epsilon s\lambda}=\frac{1}{4}\left( 1+\frac{\lambda}{F_{\epsilon s}(k_y)}(-\hbar v_Qk_y \tau_z
	+\epsilon \beta_{so}s_z\tau_z-t\tau_x) \right)\left( 1+s s_z \right).
\end{eqnarray}

Next we need to use the Green function and eigen-energy of the Hamiltonian to evaluate
$\chi^{(2)}({\bf q})$. We treat both ${\bf q}$ and the magnetic field $B_x$ (the corresponding
$Q$ and $v_Q$) as perturbations and expand superconductivity susceptibility up to
the second order in ${\bf q}$ and $B_x$. In particular, in-plane magnetic fields break the
$D_{3d}$ symmetry, and thus can couple the pairings in different representations. Direct
calculations show that the $A_{1g}$ pairing is coupled to the $A_{1u}$ pairing. 
Therefore, we first discuss the coupled $A_{1g}-A_{1u}$ pairing and then consider $E_u$ pairing. 

\subsection{$A_{1g}-A_{1u}$ pairing }
Let us label
the $A_{1g}$ pairing $\Delta_{A_{1g},1}$ and $\Delta_{A_{1g},2}$ by $\Delta_{1}$ and $\Delta_{2}$,
and the $A_{1u}$ pairing $\Delta_{A_{1u}}$ by $\Delta_3$. The corresponding Landau free
energy is given by
\begin{eqnarray}
	L=\frac{1}{2}\sum_{\bf q}\left(
	\begin{array}{ccc}
		\Delta_{1}^*&\Delta_2^*&\Delta_3^*
	\end{array}
	\right)\left(
	\begin{array}{ccc}
		\frac{1}{V_{A_{1g},1}}-\chi_{11}&-\chi_{12}&-\chi_{13}\\
		-\chi_{21}&\frac{1}{V_{A_{1g},2}}-\chi_{22}&-\chi_{23}\\
		-\chi_{31}&-\chi_{32}&\frac{1}{V_{A_{1u}}}-\chi_{33}
	\end{array}
	\right)\left(
	\begin{array}{c}
		\Delta_1\\\Delta_2\\\Delta_3
	\end{array}
	\right).
\end{eqnarray}
where $V_{A_{1g},1}=\frac{U_0}{2}, V_{A_{1g},2}=\frac{V_0}{2}$ and
$V_{A_{1u}}=\frac{U_0}{2}$. The superconductivity susceptibility in the above Landau
free energy is given by
\begin{eqnarray}
	&&\chi_{11}=4N_0\left( ln \left( \frac{2\gamma\omega_D}{\pi kT}\right) \left(
	1-\frac{(\hbar v_Qk_f)^2}{4D_0^2}\frac{t^2}{D_0^2}\right) -
	\frac{C}{2}\left( \frac{1}{2\pi kT} \right)^2\left(
	\frac{2\beta_{so}^2(\hbar v_Q k_f)^2}{D_0^2}+ \frac{(\hbar v_f q)^2}{2}\right)
	\right)\\
	&&\chi_{12}=\chi_{21}=-\frac{4tN_0}{D_0}\left( ln \left( \frac{2\gamma\omega_D}
	{\pi kT}\right) \left( 1+ \frac{\beta_{so}^2(\hbar v_Q k_f)^2}{2D_0^4} \right)
	-\frac{C}{2}\left( \frac{1}{2\pi kT} \right)^2\left( \frac{2\beta_{so}^2(\hbar v_Q
	k_f)^2}{D_0^2} + \frac{(\hbar v_fq)^2}{2} \right)\right)\\
	&&\chi_{13}=\chi_{31}=\frac{2N_0\hbar v_Qq_y}{D_0}\left( ln \left(
	\frac{2\gamma\omega_D}{\pi kT}\right) \frac{t^2}{D_0^2}
	+ \frac{2C\hbar v_f k_f\beta_{so}^2}{D_0}\left( \frac{1}{2\pi kT} \right)^2
	\right)\\
	&&\chi_{22}=4N_0\left( ln \left( \frac{2\gamma\omega_D}{\pi kT}\right) \left(
	\frac{t^2}{D_0^2}+\frac{(\hbar v_Q k_f)^2}{4D_0^2}\left( 1+\frac{(
	t^2-\beta_{so}^2)\beta_{so}^2}{D_0^4} \right)\right) - \frac{C}{2}\left(
	\frac{1}{2\pi kT} \right)^2 \frac{t^2}{D_0^2}\left(
	\frac{2\beta_{so}^2(\hbar v_Q k_f)^2}{D_0^2}+\frac{(\hbar v_f q)^2}{2}  \right)
	\right)\\
	&&\chi_{23}=\chi_{32}=\frac{2N_0 t\hbar v_Q q_y}{D_0^2}
	\left( ln \left( \frac{2\gamma\omega_D}{\pi kT}\right)
	\left( \frac{2\beta_{so}^2}{D_0^2} - 1 \right)
	-\frac{2C\hbar v_f k_f \beta_{so}^2}{D_0}\left( \frac{1}{2\pi kT} \right)^2\right)\\
        &&\chi_{33}=4N_0\left( ln \left( \frac{2\gamma\omega_D}{\pi kT}\right) \left(
	\frac{\beta_{so}^2}{D_0^2}+\frac{(\hbar v_Q k_f)^2}{4D_0^2}\left( \frac{(\beta_{so}^2
	-t^2)\beta_{so}^2}{D_0^4}-1 \right)\right) - \frac{C}{2}\left(
	\frac{1}{2\pi kT} \right)^2 \frac{\beta_{so}^2}{D_0^2}\left(
	\frac{2\beta_{so}^2(\hbar v_Q k_f)^2}{D_0^2}+\frac{(\hbar v_f q)^2}{2}  \right)
	\right).
\end{eqnarray}
The critical temperature can be obtained by minimizing the above Landau free energy, but this
is quite complicated. Therefore, we can consider the following simplifications.
We consider the limit $t\ll\beta_{so}$, in which the $\Delta_1$ pairing will dominate over
$\Delta_2$ for the $A_{1g}$ pairing. Therefore, we can substitute $\Delta_2$ by $\Delta_2=
-\frac{V_0t}{U_0D_0}\Delta_1$ and obtain the Landau free energy
\begin{eqnarray}
	L=\frac{1}{2}\sum_{\bf q}\left(
	\begin{array}{cc}
		\Delta_{1}^*&\Delta_3^*
	\end{array}
	\right)\left(
	\begin{array}{ccc}
		\frac{2}{U_0}\left( 1+\frac{V_0t^2}{U_0D_0^2} \right)
		-\tilde{\chi}_{11}&-\tilde{\chi}_{13}\\
		-\tilde{\chi}_{31}&\frac{2}{U_0}-\tilde{\chi}_{33}
	\end{array}
	\right)\left(
	\begin{array}{c}
		\Delta_1\\\Delta_3
	\end{array}
	\right),\label{App:Landau2}
\end{eqnarray}
where
\begin{eqnarray}
	&&\tilde{\chi}_{11}=\chi_{11}+\frac{V_0^2t^2}{U_0^2D_0^2}\chi_{22}
	-\frac{V_0t}{U_0D_0}(\chi_{12}+\chi_{21})=4N_0 \left( 1+\frac{V_0t^2}{U_0D_0^2} \right)^2
	ln \left( \frac{2\gamma\omega_D}{\pi kT} \right)+\mathcal{P}(v_Q,{\bf q})\\
	&&\tilde{\chi}_{13}=\tilde{\chi}_{31}=\chi_{13}-\frac{V_0t}{U_0D_0}\chi_{23}=
	\mathcal{Q}(v_Q,{\bf q})\\
	&&\tilde{\chi}_{33}=\chi_{33}=4N_0\frac{\beta_{so}^2}{D_0^2}
	ln \left( \frac{2\gamma\omega_D}{\pi kT} \right)+\mathcal{R}(v_Q,{\bf q}).
\end{eqnarray}
Here
\begin{eqnarray}
	&&\mathcal{P}=4N_0\left( ln \left( \frac{2\gamma\omega_D}{\pi kT}\right) \frac{(\hbar v_Q k_f)^2t^2}
	{4D_0^4}\left( -1 +\frac{V_0^2}{U_0^2}\left( 1+\frac{(t^2-\beta_{so}^2)\beta_{so}^2}{D_0^4}\right)
	+\frac{4V_0\beta_{so}^2}{U_0D_0^2} \right)\right.\nonumber\\
	&&\left. - \frac{C}{2}\left(\frac{1}{2\pi kT} \right)^2
	\left(\frac{2\beta_{so}^2(\hbar v_Q k_f)^2}{D_0^2}+\frac{(\hbar v_f q)^2}{2}  \right)
	\left( 1+\frac{V_0t^2}{U_0D_0^2} \right)^2\right)\\
	&&\mathcal{Q}=\frac{2N_0\hbar v_Qq_y}{D_0}\left( ln \left(
	\frac{2\gamma\omega_D}{\pi kT}\right) \frac{t^2}{D_0^2} \left( 1-\frac{V_0(\beta_{so}^2-t^2)}
	{U_0D_0^2} \right) + \frac{2C\hbar v_f k_f\beta_{so}^2}{D_0}\left( \frac{1}{2\pi kT} \right)^2
	\left( 1+\frac{V_0t^2}{U_0D_0^2} \right)\right)\\
	&&\mathcal{R}=4N_0\left( ln \left( \frac{2\gamma\omega_D}{\pi kT}\right) \frac{(\hbar v_Q k_f)^2}{4D_0^2}
	\left( \frac{(\beta_{so}^2-t^2)\beta_{so}^2}{D_0^4}-1 \right) - \frac{C}{2}\left(
	\frac{1}{2\pi kT} \right)^2 \left(\frac{2\beta_{so}^2(\hbar v_Q k_f)^2}{D_0^2}
	+\frac{(\hbar v_f q)^2}{2}  \right) \frac{\beta_{so}^2}{D_0^2}\right).
\end{eqnarray}
The corresponding linearized gap equation is given by
\begin{eqnarray}
	\left(
	\begin{array}{ccc}
		\frac{2}{U_0}\left( 1+\frac{V_0t^2}{U_0D_0^2} \right)
		-\mathcal{P}&-\mathcal{Q}\\
		-\mathcal{Q}&\frac{2}{U_0}-\mathcal{R}
	\end{array}
	\right)\left(
	\begin{array}{c}
		\Delta_1\\\Delta_3
	\end{array}
	\right)=N_0 ln\left( \frac{2\gamma\omega_D}{\pi kT} \right)
        \left(
	\begin{array}{ccc}
		\left( 1+\frac{V_0t^2}{U_0D_0^2} \right)^2
		&0\\
		0&\left( \frac{\beta_{so}^2}{D_0^2} \right)
	\end{array}
	\right)\left(
	\begin{array}{c}
		\Delta_1\\\Delta_3
	\end{array}
	\right)\label{App:gapeqn_orbital}
\end{eqnarray}
By solving this generalized eigen-equation, one can obtain two solutions
for the critical temperature $T_c$, which is a function of magnetic fields
$B_x$ and momentum ${\bf q}$. The true $T_c$ is obtained by maximizing the
larger solution with respect to ${\bf q}$. 
Fig. \ref{App_Fig:2} reveal the eigen values of Eq. (\ref{App:gapeqn_orbital}) as a function of the momentum ${\bf q}$ for different magnetic fields. 
One can clearly see that for $B_x/B_N=0.18$, the maximum $T_c$ is located at
$q_y=0$ while for $B_x/B_N=0.37$, the maximum $T_c$ is shifted to $q_y\approx 0.015 \frac{1}{nm}$. 
With this type of calculation for different magnetic fields, 
one can extract the $T_c$ and $q_c$ as a function of magnetic fields, 
as shown in Fig. 2b and c in the main text. We find a transition 
between the conventional BCS state and the FFLO state. 
Furthermore, we can substitute the maximum $T_c$ and the corresponding
eigen vector of $(\Delta_1,\Delta_3)^T$ back to the free energy (\ref{App:Landau2})
and calculate the free energy and the gap function 
as a function of magnetic fields, as shown in Fig. \ref{App_Fig:1}. 
One can see a rapid change of the slope of free energy 
at the transition point. At the same time, there is a jump 
in the gap function. These features indicate a first-order transition
in the current case. This transition will be further discussed
in details below. 

\begin{figure} 
\includegraphics[width=15cm]{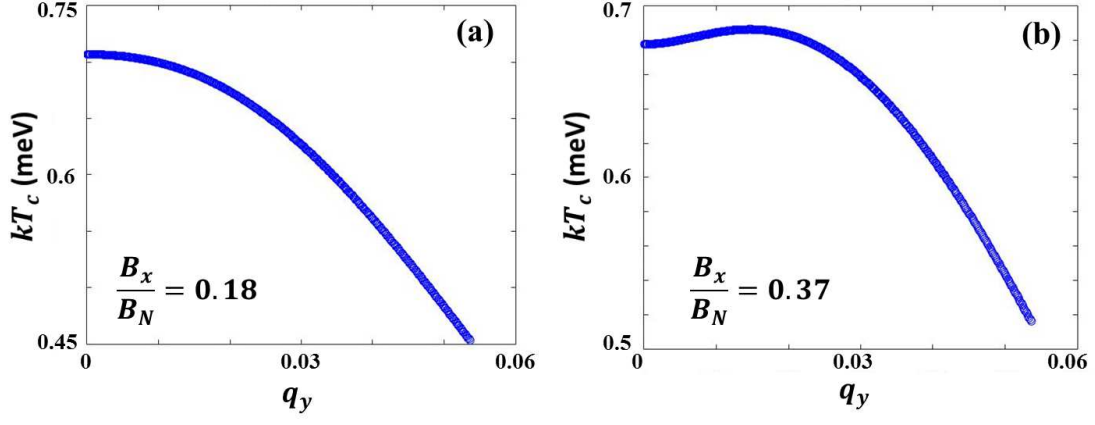}
\caption{The critical temperature as a function of momentum $q_y$ for (a) $B_x/B_N=0.18$
and (b) $B_x/B_N=0.37$. $B_N=\frac{2kT_{c0}}{v_fz_0}$.}
\label{App_Fig:2}
\end{figure}

\begin{figure} 
\includegraphics[width=15cm]{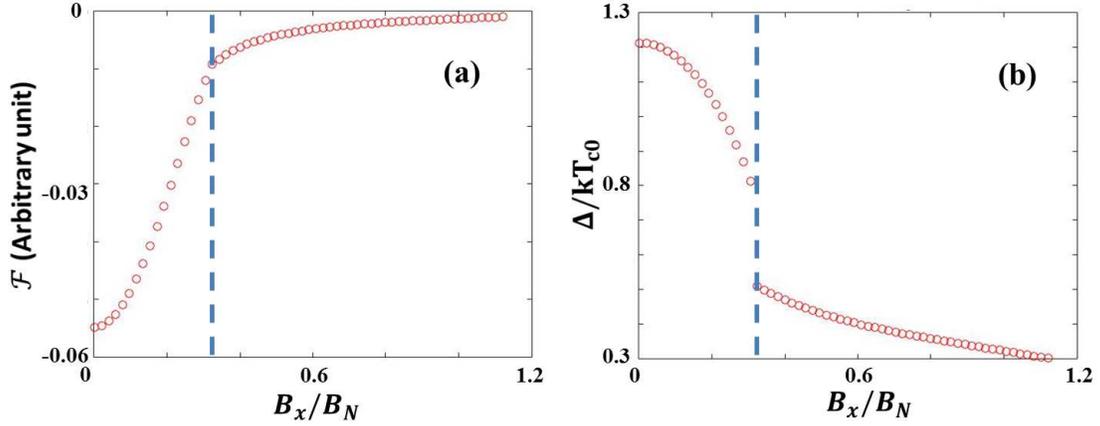}
\caption{(a) Free energy as a function of magnetic fields. 
(b) Gap function as a function of magnetic fields. 
Here the dashed blue lines are for the phase transition point.
$B_N=\frac{2kT_{c0}}{v_fz_0}$. }
\label{App_Fig:1}
\end{figure}


\subsection{The decoupling limit $t\rightarrow 0$}
In this section, we first consider a simple
limit $t\rightarrow 0$, in which we find $D_0=\beta_{so}$ and
$T_{c0,A_{1g}}=T_{c0,A_{1u}}=\frac{2\gamma\omega_D}{\pi k} exp\left(-1/(2U_0N_0)\right)=T_{c0}$
and $\mathcal{P}=\mathcal{R}$.
In this case, the Landau free energy takes a simple form
\begin{eqnarray}
	L=\frac{1}{2}\sum_{\bf q}\left(
	\begin{array}{cc}
		\Delta_{1}^*&\Delta_3^*
	\end{array}
	\right)\left(
	\begin{array}{ccc}
		4N_0 ln\left( \frac{T}{T_{c0}} \right)-\mathcal{P}(v_Q,{\bf q})&
		-\mathcal{Q}(v_Q,{\bf q})\\-\mathcal{Q}(v_Q,{\bf q})&
		4N_0 ln\left( \frac{T}{T_{c0}} \right)-\mathcal{P}(v_Q,{\bf q})
	\end{array}
	\right)\left(
	\begin{array}{c}
		\Delta_1\\\Delta_3
	\end{array}
	\right),
\end{eqnarray}
where
\begin{eqnarray}
	&&\mathcal{P}(v_Q,{\bf q})=-N_0C\left( \frac{1}{2\pi kT} \right)^2
	\left( 4(\hbar v_Q k_f)^2+(\hbar v_f q)^2\right)\\
	&&\mathcal{Q}(v_Q,{\bf q})=4N_0C(\hbar v_f k_f) \left( \frac{1}{2\pi kT} \right)^2
	(\hbar v_Q q_y)
\end{eqnarray}
Since $v_Q\propto B_x$, we find that $\Delta_1$ and $\Delta_3$ pairings are coupled to each
other by a new term with the form $\Delta_1^*B_x q_y\Delta_3$.We notice a similar term
$\Delta^* B_iq_j\Delta$ ($i,j$ are two indices for axis) describes the electro-magnetic effect
in non-centrosymmetric superconductors. Since $\Delta_1$ and $\Delta_3$
have opposite parities under inversion, this term does not break inversion symmetry,
in sharp contrast to the term $\Delta^* B_iq_j\Delta$ in non-centrosymmetric superconductors.

According to the above Landau free energy, 
the critical temperature can be obtained by
maximizing the following function
\begin{eqnarray}
	4N_0 ln \left( \frac{T_c}{T_{c0}}\right)=\mathcal{P}+ |\mathcal{Q}|.
\end{eqnarray}
By substituting the form of $\mathcal{P}$ and $\mathcal{Q}$,
we can re-write the above equation as
\begin{eqnarray}
	ln \left( \frac{T_c}{T_{c0}}\right)=-\frac{C}{4}\left( \frac{1}{2\pi kT_{c0}} \right)^2
	\left( (\hbar v_f q_x)^2+(\hbar v_f |q_y|- 2\hbar |v_Q| k_f)^2 \right).
\end{eqnarray}
It is easy to see that when we choose $q_x=0$ and $|q_y|=\frac{2 k_f |v_Q|}{v_f}$, the right
hand side of the above expression is maximized. The corresponding $T_c$ is given by
\begin{eqnarray}
	ln \left( \frac{T_c}{T_{c0}}\right)= 0,
\end{eqnarray}
from which one can see there is no correction to $T_c$. In contrast, for ${\bf q}=0$, we find
\begin{eqnarray}
	ln \left( \frac{T_c({\bf q=0})}{T_{c0}}\right)=-C\left( \frac{\beta_{so}}{2\pi kT_{c0}} \right)^2
	 \frac{(\hbar v_Q k_f)^2}{\beta_{so}^2},
\end{eqnarray}
which is always smaller than the finite momentum pairing.
Therefore, we conclude that under in-plane magnetic fields,
the stable superconducting phase occurs for a non-zero $q_y$,
leading to the FFLO state.
The stable $q_y$ is given by $|q_y|=q_c=\frac{2 k_f |v_Q|}{v_f}
=\frac{2 k_f m}{\hbar k_f}\frac{eB_xz_0}{2m}=\frac{eB_xz_0}{\hbar}$.
With flux quantum $\phi_0=\frac{h}{2e}$, we have $\lambda_c=\frac{2\pi}{q_c}
=\frac{2\pi\hbar }{eB_x z_0}=\frac{2\phi_0}{B_x z_0}$.
Thus, the wave length of the FFLO state is determined by 
the corresponding area for magnetic flux quantum.

We notice that $q_y=q_c$ and $q_y=-q_c$ are two degenerate states. To see this, we perform
a transformation $\Delta_+=\frac{1}{\sqrt{2}}\left( \Delta_1+\Delta_3 \right)$
and $\Delta_-=\frac{1}{\sqrt{2}}\left( \Delta_1-\Delta_3 \right)$ and the corresponding
Landau free energy is transformed as
\begin{eqnarray}
	L=\frac{1}{2}\sum_{\bf q}\left(
	\begin{array}{cc}
		\Delta_{+}^*&\Delta_-^*
	\end{array}
	\right)\left(
	\begin{array}{ccc}
		4N_0 ln\left( \frac{T}{T_{c0}} \right)-\mathcal{P}(v_Q,{\bf q})
		-\mathcal{Q}(v_Q,{\bf q})&0\\0&
		4N_0 ln\left( \frac{T}{T_{c0}} \right)-\mathcal{P}(v_Q,{\bf q})
		+\mathcal{Q}(v_Q,{\bf q})
	\end{array}
	\right)\left(
	\begin{array}{c}
		\Delta_+\\\Delta_-
	\end{array}
	\right).
\end{eqnarray}
Physically, $\Delta_+$ describes the pairing in the top layer and $\Delta_-$ is for the pairing in the bottom layer. Thus, the diagonal form of the 
above Landau free energy just corresponds to the decoupling between 
two layers. 

Let's assume $B_x>0$ and if $q_y>0$, $\mathcal{Q}>0$ and thus $\Delta_+$ is favored.
For $q_y<0$, $\mathcal{Q}<0$ and correspondingly, $\Delta_-$ is favored. 
Since $\Delta_+(q_c)$ and $\Delta_-(-q_c)$ pairings are degenerate,
the full real space expression of the FFLO state is given by
\begin{eqnarray}
	\Delta({\bf r})=\Delta_+(q_c)e^{iq_cy}+\Delta_-(-q_c)e^{-iq_cy}.
\end{eqnarray}
In the above expression, the first term describes the helical phase
in the top layer while the second term is for the helical phase in the
bottom layer with opposite momentum. 

The relative magnitude of $\Delta_+(q_c)$ and $\Delta_-(-q_c)$ are
determined by the fourth order term in Landau free enregy. 
The general form of the fourth order term is 
\begin{eqnarray}
	L_4=\mathcal{B}_s\left( |\Delta_+(q_c)|^2+|\Delta_-(-q_c)|^2 \right)^2
	+\mathcal{B}_a\left( |\Delta_+(q_c)|^2-|\Delta_-(-q_c)|^2 \right)^2.
	\label{Eq:App_L4_qc}
\end{eqnarray}
If $\mathcal{B}_a>0$, we need $|\Delta_+|=|\Delta_-|$ to minimize the second term in
the above expression. This corresponds to the stripe phase with its pairing amplitude
oscillating in the real space. If $\mathcal{B}_s<0$, we have either $|\Delta_+|=0$ or
$|\Delta_-|=0$, which corresponds to the helical phase, in which only the phase oscillates
while the amplitude persists.

Microscopically, the fourth order term can be computed from 
(\ref{eq:APP_fourth}). 
More specifically, the fourth order terms for $A_{1g}$ and $A_{1u}$
pairings are given by
\begin{eqnarray}
	&&L_{4,A_{1g}}=\frac{7N_0\zeta(3)}{16(\pi kT_{c0})^2}
	\sum_{\left\{ q \right\}}\Delta_{1}^*(q_1)\Delta_1^*(q_3)
	\Delta_1(q_2)\Delta_1(q_4)\delta_{q_1+q_3=q_2+q_4}\\
	&&L_{4,A_{1u}}=\frac{7N_0\zeta(3)}{16(\pi kT_{c0})^2}
	\frac{\beta_{so}^4}{D_0^4}
	\sum_{\left\{ q \right\}}\Delta_{3}^*(q_1)\Delta_3^*(q_3)
	\Delta_3(q_2)\Delta_3(q_4)\delta_{q_1+q_3=q_2+q_4}, 
\end{eqnarray}
where the summation $\{q\}$ is for $q_1, q_2, q_3, q_4$. 
For the study of the FFLO state, it turns out that one also needs
to take into account the coupling between
the $A_{1g}$ and $A_{1u}$ pairing for the fourth order term, which is given by
\begin{eqnarray}
	&&L_{4,A_{1g}-A_{1u}}=\frac{7N_0\zeta(3)}{16(\pi kT_{c0})^2}
	\frac{\beta_{so}^2}{D_0^2}
	\sum_{\left\{ q \right\}}\left( \Delta_{1}^*(q_1)\Delta_3^*(q_3)
	\Delta_1(q_2)\Delta_3(q_4) + \Delta_{3}^*(q_1)\Delta_1^*(q_3)
	\Delta_3(q_2)\Delta_1(q_4) + \Delta_{1}^*(q_1)\Delta_1^*(q_3)
	\Delta_3(q_2)\Delta_3(q_4)\right. \nonumber\\
	&&\left. + \Delta_{3}^*(q_1)\Delta_3^*(q_3)
	\Delta_1(q_2)\Delta_1(q_4) + \Delta_{3}^*(q_1)\Delta_1^*(q_3)
	\Delta_1(q_2)\Delta_3(q_4) + \Delta_{1}^*(q_1)\Delta_3^*(q_3)
	\Delta_3(q_2)\Delta_1(q_4)\right)\delta_{q_1+q_3=q_2+q_4}
\end{eqnarray}

Collecting all the above terms, we find the fourth order term for 
the $A_{1g}$ and $A_{1u}$ pairings can be written as a compact form
\begin{eqnarray}
	&&L_{4}=\frac{7N_0\zeta(3)}{32(\pi kT_{c0})^2}\sum_{\{q\},\alpha=\pm}
	\left( \Delta_1^*(q_1)+\alpha \frac{\beta_{so}}{D_0}\Delta_3^*(q_1)
	\right) \left( \Delta_1^*(q_3)+\alpha \frac{\beta_{so}}{D_0}
	\Delta_3^*(q_3)\right) \nonumber\\
	&&\left( \Delta_1(q_2)+\alpha 
	\frac{\beta_{so}}{D_0}\Delta_3(q_2)\right)
	\left( \Delta_1(q_4)+\alpha \frac{\beta_{so}}{D_0}\Delta_3(q_4)\right)
	\delta_{q_1+q_3=q_2+q_4}. \label{APP:eq_fourthA1gA1u}
\end{eqnarray}
In the limit $t\rightarrow 0$, the fourth order term is reduced to the 
form
\begin{eqnarray}
	L_4=\frac{7N_0\zeta(3)}{8(\pi kT_{c0})^2}\sum_{\{q\},\alpha=\pm}
	\Delta_{\alpha}^*(q_1)\Delta_{\alpha}^*(q_3)
	\Delta_{\alpha}(q_2)\Delta_{\alpha}(q_4)\delta_{q_1+q_3=q_2+q_4},
\end{eqnarray}
which is also decoupled between the top and bottom
layers. In combining with the form of $L_2$ in Eq. (6) in the main text,
we conclude that the Landau free energy is decoupled between
the top and bottom layers for the limit $t\rightarrow 0$. 

We consider the fourth order term for $\Delta_\pm(\pm q_c)$, which 
takes the form
\begin{eqnarray}
	L_4=\frac{7N_0\zeta(3)}{8(\pi kT_{c0})^2}\sum_{\alpha=\pm}
        \left( |\Delta_\alpha(q_c)|^4+|\Delta_\alpha(-q_c)|^4
	+4|\Delta_\alpha(q_c)|^2|\Delta_\alpha(-q_c)|^2\right)
\end{eqnarray}
Since only $\Delta_+(q_c)$ and $\Delta_-(-q_c)$ are favored by the second order
term, the fourth order term involving $\Delta_+(q_c)$ and $\Delta_-(-q_c)$ is given by
\begin{eqnarray}
	L_4=\frac{7N_0\zeta(3)}{16(\pi kT_{c0})^2}
	\left( \left( |\Delta_+(q_c)|^2+|\Delta_-(-q_c)|^2 \right)^2
	+ \left( |\Delta_+(q_c)|^2-|\Delta_-(-q_c)|^2 \right)^2\right).
\end{eqnarray}

Compared with Eq. (\ref{Eq:App_L4_qc}), we find $\mathcal{B}_s=
\mathcal{B}_a= \frac{7N_0\zeta(3)}{16(\pi kT_{c0})^2}$.
Since $\mathcal{B}_a>0$, we conclude that stripe phase will be favored when
the temperature is close to $T_c$.

\subsection{Phase transition between the BCS superconducting state 
and the FFLO state}
The phase diagram of the gap function as a function of magnetic fields and temperature 
is shown in the main text. Here we will present more analytical results
and show that the transition between the uniform $A_{1g}$ pairing 
and the FFLO state is of the first order. 

By solving the gap equation (\ref{App:gapeqn_orbital}), we can obtain 
the $T_c$ as a function of momentum ${\bf q}$ and magnetic field $B_x$. 
Since the state with $q_x=0$ will always be favored, we focus on $q_y$ here. 
The critical temperature
is an even function of $q_y$ and it can be expanded as
$T_c(q_y)=T_{c0}+T_{c1}q_y^2+T_{c2}q_y^4$ up to the fourth order
in $q_y$, where $T_{c2}<0$. The transition between uniform superconductivity
and FFLO state occurs when $T_{c1}$ changes from negative to positive. 
Since the transition is tuned by magnetic field $B_x$, the parameter 
$T_{c1}$ takes the form $T_{c1}=S_1(B_x-B_{c0})$, where $S_1>0$ 
and $B_{c0}$ labels the critical magnetic field. 
For a postive $T_{c1}$, the maximum $T_c$ is achieved
when $q_y=\pm\sqrt{\frac{S_1}{2T_{c2}}(B_{c0}-B_x)}$
and the corresponding $T_c$ is $T_c(q_y)=T_{c0}-\frac{S_1^2}{4T_{c2}}
(B_{c0}-B_x)^2$. 
In addition, $T_{c0}$ should be an even function of magnetic field, 
$T_{c0}=R_1-R_2B_x^2$. 
We focus on the magnetic field around $B_{c0}$ and thus expand $T_{c0}$
as $T_{c0}=\tilde{T}_{c0}-2R_2B_{c0}\delta B$ with $\tilde{T}_{c0}
=R_1-R_2B_{c0}^2$. Furthermore, the eigen-vector for 
the corresponding $T_c$ in Eq. (\ref{App:gapeqn_orbital}) 
is simplified as 
\begin{eqnarray}
	\left(
	\begin{array}{c}
		\Delta_1\\\Delta_3
	\end{array}
	\right)=\Delta_{0q}\left(
	\begin{array}{c}
		1\\0
	\end{array}
	\right)
\end{eqnarray}
around $B_{c0}$ since the $A_{1g}$ pairing will always dominate
and the amplitude $\Delta_{0q}$ is to 
be determined. With these simplifications, we are able
to evaluate the free energy analytically. The second order term is 
derived as
\begin{eqnarray}
	L_2=2N_0\sum_{q_y}|\Delta_{0q}|^2 ln\left( \frac{T}{T_{c}(q)} \right)
	\left( 1+\frac{V_0t^2}{U_0D_0^2} \right)^2. 
\end{eqnarray}
The main difference between zero momentum and finite momentum pairings lies
in the fourth order term. 

For zero momentum pairing ($T_{c1}<0$), the fourth order term is
given by
\begin{eqnarray}
	L_4=\frac{7N_0\zeta(3)}{16(\pi kT_{c0})^2}|\Delta_{00}|^4. 
\end{eqnarray}
By minimizing $L_2+L_4$ for $q_y=0$, we find the minimal free energy is given by
\begin{eqnarray}
	\mathcal{F}_{q_y=0}=-\frac{16N_0(\pi kT_{c0})^2}{7\zeta(3)}
	\left( ln\left( \frac{T}{T_{c0}} \right) \right)^2
	\left( 1+\frac{V_0t^2}{U_0D_0^2} \right)^4
\end{eqnarray}
as a function temperature $T$, which is assumed to be close to $T_{c0}$. 

On the other hand, for a non-zero $q_y$, the fourth order term is much
more complicated. The $q_1, q_2, q_3$ and $q_4$ in Eq. 
(\ref{APP:eq_fourthA1gA1u}) can take the following six cases: (1) 
$q_1=q_2=q_3=q_4=q_y$; (2) $q_1=q_2=q_3=q_4=-q_y$; (3)
$q_1=q_2=-q_3=-q_4=q_y$; (4) $q_1=-q_2=-q_3=q_4=q_y$; (5)
$-q_1=-q_2=q_3=q_4=q_y$; (6) $-q_1=q_2=-q_3=-q_4=q_y$.
Furthermore, we only focus on the stripe phase, which has been 
confirmed in numerical calculations. Thus, we take 
$|\Delta_{0,-q}|=|\Delta_{0q}|$ and as a result, 
the full Landau free energy takes the form 
\begin{eqnarray}
	L= 4N_0|\Delta_{0q}|^2 ln\left( \frac{T}{T_{c}(q)} \right)
	\left( 1+\frac{V_0t^2}{U_0D_0^2} \right)^2+
	\frac{21N_0\zeta(3)}{8(\pi kT_{c}(q_y))^2}|\Delta_{0q}|^4, 
\end{eqnarray}
the minimal of which gives the temperature dependence of 
the free energy
\begin{eqnarray}
	\mathcal{F}_{q_y\neq 0}=-\frac{32N_0(\pi kT_{c}(q_y))^2}{21\zeta(3)}
	\left( ln\left( \frac{T}{T_{c}(q_y)} \right) \right)^2
	\left( 1+\frac{V_0t^2}{U_0D_0^2} \right)^4. 
\end{eqnarray}

Now let's fix the temperature $T$ smaller than $T_{c0}$ and study 
the phase transition by varying magnetic field $B_x$. 
The transition happens at $B_{c0}$ for the temperature at $T_{c0}$
but will shift a bit away when the temperature $T$ is lower than 
$T_{c0}$. To see that, we need to consider the limit of $q_y\rightarrow 0$. 
In this limit, $T_c(q_y)\rightarrow T_{c0}$ and thus
\begin{eqnarray}
	\mathcal{F}_{q_y\rightarrow 0}=-\frac{32N_0(\pi kT_{c0})^2}{21\zeta(3)}
	\left( ln\left( \frac{T}{T_{c0}} \right) \right)^2
	\left( 1+\frac{V_0t^2}{U_0D_0^2} \right)^4
	=\frac{2}{3}\mathcal{F}_{q_y=0}>\mathcal{F}_{q_y=0}
\end{eqnarray}
Thus, at $B_x=B_{c0}$, the zero momentum pairing is more stable. 
The difference comes from the form of fourth order term when 
$q_y=0$ and $q_y\neq0$. 
To determine the critical magnetic field at $T$, we may expand
the free energy around $B_{c0}$. With $B_x=B_{c0}+\delta B$, we find
\begin{eqnarray}
	F_{q_y=0}=-\frac{16N_0(\pi k)^2}{7\zeta(3)}
	\left( 1+\frac{V_0t^2}{U_0D_0^2} \right)^4
	\left( \left( ln\left( \frac{T}{\tilde{T}_{c0}} \right) \right)^2
	\tilde{T}_{c0}^2+4R_2B_{c0}\tilde{T}_{c0} ln 
	\left( \frac{T}{\tilde{T}_{c0}} \right) \left( 1- ln 
	\left( \frac{T}{\tilde{T}_{c0}} \right) \right)\delta B
 \right)
\end{eqnarray}
and 
\begin{eqnarray}
	F_{q_y\rightarrow 0}=-\frac{32N_0(\pi k)^2}{21\zeta(3)}
	\left( 1+\frac{V_0t^2}{U_0D_0^2} \right)^4
	\left( \left( ln\left( \frac{T}{\tilde{T}_{c0}} \right) \right)^2
	\tilde{T}_{c0}^2+4R_2B_{c0}\tilde{T}_{c0} ln 
	\left( \frac{T}{\tilde{T}_{c0}} \right) \left( 1- ln 
	\left( \frac{T}{\tilde{T}_{c0}} \right) \right)\delta B
 \right)
\end{eqnarray}
up to the first order in $\delta B$. 
Thus, the critical magnetic field is determined by 
\begin{eqnarray}
	\delta B_c=- \frac{ ln \left( T/\tilde{T}_{c0} \right)\tilde{T}_{c0}}
	{4R_2B_{c0}\left( 1- ln (T/\tilde{T}_{c0}) \right)}
\end{eqnarray}
when $T\rightarrow \tilde{T}_{c0}^- $ and $ln 
\left( \frac{T}{\tilde{T}_{c0}} \right)$ is treated as a small number. 

At the $\delta B_c$, the first derivative of the free energy is given by
\begin{eqnarray}
	\frac{\partial F_{q_y=0}}{\partial \delta B}=	
	-\frac{16N_0(\pi k)^2}{7\zeta(3)}
	\left( 1+\frac{V_0t^2}{U_0D_0^2} \right)^4
	4R_2B_{c0}\tilde{T}_{c0} ln 
	\left( \frac{T}{\tilde{T}_{c0}} \right)
\end{eqnarray}
and 
\begin{eqnarray}
	\frac{\partial F_{q_y\rightarrow 0}}{\partial \delta B}=	
	-\frac{32N_0(\pi k)^2}{21\zeta(3)}
	\left( 1+\frac{V_0t^2}{U_0D_0^2} \right)^4
	4R_2B_{c0}\tilde{T}_{c0} ln 
	\left( \frac{T}{\tilde{T}_{c0}} \right)
\end{eqnarray}
up to the first order in $ln 
\left( \frac{T}{\tilde{T}_{c0}} \right)$. 
We find $\frac{\partial F_{q_y=0}}{\partial \delta B}\neq
\frac{\partial F_{q_y\rightarrow 0}}{\partial \delta B}$ 
at $\delta B_c$, thus confirming the phase transition 
is of the first order nature. 

\subsection{FFLO/BCS Josephson junction}
Next we will discuss the possible detection of the FFLO state. 
A Josephson junction between the FFLO state and a conventional superconductor
is considered, as shown in Fig. 1d in the main text. 
The Josephson current in this system is given by 
\begin{eqnarray}
	I_J=Im\left( t \int dy \Psi^*_{BCS}(y)\Psi_{FFLO)(y)} \right). 
	\label{App:Eq_Josephson}
\end{eqnarray}
where $\Psi_{FFLO}$ and $\Psi_{BCS}$ are the pair functions for
the FFLO state and conventional superconductors. $t$ is the hopping parameter. 
We take the form $\Psi_{BCS}=\Psi_0 e^{i\varphi}$, where $\varphi$ is
the phase factor, and $\Psi_{FFLO}=\Psi_+e^{iq_cy}+\Psi_-e^{-iq_cy}$
according to the form of the gap function. With these forms of the pair
functions, the Josephson current is found to be 
\begin{eqnarray}
	I_J=\sum_{\alpha=\pm} I_{c,\alpha} \frac{\sin (\alpha q_cL/2)}
	{(\alpha q_cL/2)} \sin \varphi
	\label{App:Eq_Josephson1}
\end{eqnarray}
where the y-direction integral is taken from $-L/2$ to $L/2$ 
and thus the maximum supercurrent is 
\begin{eqnarray}
	I_m=\left|\sum_{\alpha=\pm} I_{c,\alpha} \frac{\sin (\alpha q_cL/2)}
	{(\alpha q_cL/2)} \right|. 
\end{eqnarray}

The maximum supercurrent will depend on the details of Josephson contact, and
if we assume $I_{c,+}=I_{c,-}=I_{c0}$, we find 
\begin{eqnarray}
	I_m=2I_{c0}\frac{\sin (q_cL/2)}{(q_cL/2)}.
\end{eqnarray}
Since $q_c\propto B_x$, the maximum supercurrent reveals an interference pattern
for in-plane magnetic fields, just like the Fraunhofer pattern in a normal
Josephson junction under out-of-plane magnetic fields \cite{kaur2005}. 

For a large in-plane magnetic field with $q_cL\ll 1$, $I_m$ will decrease to zero, but an additional magnetic field along the surface normal will lead to an asymmetric Fraunhofer pattern \cite{yang2000}.

\subsection{$E_u$ pairing}
Finally we discuss the orbital effect of the $E_u$ pairing. 
Direct calculation shows that up to the
second order in $v_Q$ and ${\bf q}$,
\begin{eqnarray}
	\chi_{E_u,11}=\chi_{E_u,22}=2N_0\left( ln \left( \frac{2\gamma\omega_D}{\pi kT}
	\right) \left( \frac{2\beta_{so}^2}{D_0^2}+\frac{t^4-5\beta_{so}^2t^2}
	{2D_0^6}(\hbar v_Q k_f)^2 \right)-\frac{C\beta_{so}^2}{D_0^2}\left( \frac{1}{2\pi kT}
	\right)^2\left( \frac{(\hbar v_f )^2q^2}{2}+\frac{\beta_{so}^2(\hbar v_Q q_y)^2}{D_0^2}
	\right)	\right)
\end{eqnarray}
and
\begin{eqnarray}
	\chi_{E_u,12}=\chi_{E_u,21}=0.
\end{eqnarray}

As a consequence, the corresponding Landau free energy still follows the standard form
with the $T_c$ given by
\begin{eqnarray}
	ln \left( \frac{T_{c,E_u}}{T_{c0,E_u}} \right)=\frac{t^4-5\beta_{so}^2t^2}
	{4\beta_{so}^2D_0^4} ln \left( \frac{2\gamma\omega_D}{\pi kT_{c0,E_u}} \right)
	(\hbar v_Q k_f)^2.
\end{eqnarray}

\section{The orbital effect of out-of-plane magnetic fields}
Finally, we would like to comment about the orbital effect of out-of-plane magnetic fields.
We consider the form of Landau free energy in the real space as
\begin{eqnarray}
	L=\int d{\bf r} \left( \mathcal{K} |\nabla \Delta ({\bf r})|^2 +\mathcal{A}
	|\Delta({\bf r})|^2 \right)
\end{eqnarray}
where $\mathcal{A}=\mathcal{C} ln \left( \frac{T}{T_{c0}} \right)$.
The orbital effect of magnetic fields is taken into account by
$-i\nabla \rightarrow \mathfrak{D}=-i\nabla + \frac{2e}{\hbar }{\bf A}$,
where the vector potential ${\bf A}$ can be chosen as ${\bf A}=(0,B_zx,0)$.
The corresponding linearized gap equation is given by
\begin{eqnarray}
	\left( \mathcal{K}\mathfrak{D}^2 + \mathcal{A} \right)\Delta({\bf r}) = 0.
\end{eqnarray}

This is nothing but the Landau level problem, which can be solved exactly by introducing
the boson operators
\begin{eqnarray}
	a=\sqrt{\frac{\hbar}{4eh_z}}(\pi_x-i\pi_y),\nonumber\\
	a^{\dag}=\sqrt{\frac{\hbar}{4eh_z}}(\pi_x+i\pi_y),
\end{eqnarray}
where $\pi_{x(y)}=-i\partial_{x(y)}+\frac{2e}{\hbar}A_{x(y)}$.
The above gap equation can be simplified as
\begin{eqnarray}
	\left( \frac{4eB_z}{\hbar}\mathcal{K}(a^{\dag}a+\frac{1}{2}) + \mathcal{A} \right)
	=0.
\end{eqnarray}
The lowest eigen-energy of the above equation is
\begin{eqnarray}
	\frac{2eB_z}{\hbar}\mathcal{K}+\mathcal{A}=0,
\end{eqnarray}
leading to the correction
\begin{eqnarray}
	ln \left( \frac{T_c}{T_{c0}} \right)=-\frac{2e}{\hbar}\frac{\mathcal{K}}{\mathcal{C}}B_z
\end{eqnarray}
which is linear in $B_z$.Therefore, the $T_c$ correction from the out-of-plane magnetic fields
is determined by the ratio $\frac{\mathcal{K}}{\mathcal{C}}$.
By looking at the parameters in Landau free energy, we find for the representation $i$ ($i=A_{1g}, A_{1u}, E_u$),
the ratio is given by $\frac{\mathcal{K}_i}{\mathcal{C}_i}=\frac{C}{6}\left( \frac{\hbar v_f}{2\pi kT_{c0,i}}
\right)^2$. Therefore, the correction is determined by $T_{c0,i}$, and a weaker $T_{c,i}$ correction
for the pairing with higher $T_{c0,i}$. The stable superconducting phase will always be stable
under an out-of-plane magnetic field. 

\end{widetext}


\begin{thebibliography}{44}%
\makeatletter
\providecommand \@ifxundefined [1]{%
 \@ifx{#1\undefined}
}%
\providecommand \@ifnum [1]{%
 \ifnum #1\expandafter \@firstoftwo
 \else \expandafter \@secondoftwo
 \fi
}%
\providecommand \@ifx [1]{%
 \ifx #1\expandafter \@firstoftwo
 \else \expandafter \@secondoftwo
 \fi
}%
\providecommand \natexlab [1]{#1}%
\providecommand \enquote  [1]{``#1''}%
\providecommand \bibnamefont  [1]{#1}%
\providecommand \bibfnamefont [1]{#1}%
\providecommand \citenamefont [1]{#1}%
\providecommand \href@noop [0]{\@secondoftwo}%
\providecommand \href [0]{\begingroup \@sanitize@url \@href}%
\providecommand \@href[1]{\@@startlink{#1}\@@href}%
\providecommand \@@href[1]{\endgroup#1\@@endlink}%
\providecommand \@sanitize@url [0]{\catcode `\\12\catcode `\$12\catcode
  `\&12\catcode `\#12\catcode `\^12\catcode `\_12\catcode `\%12\relax}%
\providecommand \@@startlink[1]{}%
\providecommand \@@endlink[0]{}%
\providecommand \url  [0]{\begingroup\@sanitize@url \@url }%
\providecommand \@url [1]{\endgroup\@href {#1}{\urlprefix }}%
\providecommand \urlprefix  [0]{URL }%
\providecommand \Eprint [0]{\href }%
\providecommand \doibase [0]{http://dx.doi.org/}%
\providecommand \selectlanguage [0]{\@gobble}%
\providecommand \bibinfo  [0]{\@secondoftwo}%
\providecommand \bibfield  [0]{\@secondoftwo}%
\providecommand \translation [1]{[#1]}%
\providecommand \BibitemOpen [0]{}%
\providecommand \bibitemStop [0]{}%
\providecommand \bibitemNoStop [0]{.\EOS\space}%
\providecommand \EOS [0]{\spacefactor3000\relax}%
\providecommand \BibitemShut  [1]{\csname bibitem#1\endcsname}%
\let\auto@bib@innerbib\@empty
\bibitem [{\citenamefont {Sigrist}\ and\ \citenamefont
  {Ueda}(1991)}]{sigrist1991a}%
  \BibitemOpen
  \bibfield  {author} {\bibinfo {author} {\bibfnamefont {M.}~\bibnamefont
  {Sigrist}}\ and\ \bibinfo {author} {\bibfnamefont {K.}~\bibnamefont {Ueda}},\
  }\href@noop {} {\bibfield  {journal} {\bibinfo  {journal} {Reviews of Modern
  physics}\ }\textbf {\bibinfo {volume} {63}},\ \bibinfo {pages} {239}
  (\bibinfo {year} {1991})}\BibitemShut {NoStop}%
\bibitem [{\citenamefont {Mineev}\ and\ \citenamefont
  {Samokhin}(1999)}]{mineev1999a}%
  \BibitemOpen
  \bibfield  {author} {\bibinfo {author} {\bibfnamefont {V.~P.}\ \bibnamefont
  {Mineev}}\ and\ \bibinfo {author} {\bibfnamefont {K.}~\bibnamefont
  {Samokhin}},\ }\href@noop {} {\emph {\bibinfo {title} {Introduction to
  unconventional superconductivity}}}\ (\bibinfo  {publisher} {CRC Press},\
  \bibinfo {year} {1999})\BibitemShut {NoStop}%
\bibitem [{\citenamefont {Bauer}\ and\ \citenamefont
  {Sigrist}(2012)}]{bauer2012}%
  \BibitemOpen
  \bibfield  {author} {\bibinfo {author} {\bibfnamefont {E.}~\bibnamefont
  {Bauer}}\ and\ \bibinfo {author} {\bibfnamefont {M.}~\bibnamefont
  {Sigrist}},\ }\href@noop {} {\emph {\bibinfo {title} {Non-centrosymmetric
  superconductors: introduction and overview}}},\ Vol.\ \bibinfo {volume}
  {847}\ (\bibinfo  {publisher} {Springer Science \& Business Media},\ \bibinfo
  {year} {2012})\BibitemShut {NoStop}%
\bibitem [{\citenamefont {Dimitrova}\ and\ \citenamefont
  {Feigel'man}(2003)}]{dimitrova2003a}%
  \BibitemOpen
  \bibfield  {author} {\bibinfo {author} {\bibfnamefont {O.~V.}\ \bibnamefont
  {Dimitrova}}\ and\ \bibinfo {author} {\bibfnamefont {M.~V.}\ \bibnamefont
  {Feigel'man}},\ }\href@noop {} {\bibfield  {journal} {\bibinfo  {journal}
  {Journal of Experimental and Theoretical Physics Letters}\ }\textbf {\bibinfo
  {volume} {78}},\ \bibinfo {pages} {637} (\bibinfo {year} {2003})}\BibitemShut
  {NoStop}%
\bibitem [{\citenamefont {Agterberg}(2003)}]{agterberg2003a}%
  \BibitemOpen
  \bibfield  {author} {\bibinfo {author} {\bibfnamefont {D.}~\bibnamefont
  {Agterberg}},\ }\href@noop {} {\bibfield  {journal} {\bibinfo  {journal}
  {Physica C: Superconductivity}\ }\textbf {\bibinfo {volume} {387}},\ \bibinfo
  {pages} {13} (\bibinfo {year} {2003})}\BibitemShut {NoStop}%
\bibitem [{\citenamefont {Barzykin}\ and\ \citenamefont
  {Gor'kov}(2002)}]{barzykin2002}%
  \BibitemOpen
  \bibfield  {author} {\bibinfo {author} {\bibfnamefont {V.}~\bibnamefont
  {Barzykin}}\ and\ \bibinfo {author} {\bibfnamefont {L.~P.}\ \bibnamefont
  {Gor'kov}},\ }\href@noop {} {\bibfield  {journal} {\bibinfo  {journal}
  {Physical review letters}\ }\textbf {\bibinfo {volume} {89}},\ \bibinfo
  {pages} {227002} (\bibinfo {year} {2002})}\BibitemShut {NoStop}%
\bibitem [{\citenamefont {Aoyama}\ and\ \citenamefont
  {Sigrist}(2012)}]{aoyama2012}%
  \BibitemOpen
  \bibfield  {author} {\bibinfo {author} {\bibfnamefont {K.}~\bibnamefont
  {Aoyama}}\ and\ \bibinfo {author} {\bibfnamefont {M.}~\bibnamefont
  {Sigrist}},\ }\href@noop {} {\bibfield  {journal} {\bibinfo  {journal}
  {Physical review letters}\ }\textbf {\bibinfo {volume} {109}},\ \bibinfo
  {pages} {237007} (\bibinfo {year} {2012})}\BibitemShut {NoStop}%
\bibitem [{\citenamefont {Yoshida}\ \emph {et~al.}(2013)\citenamefont
  {Yoshida}, \citenamefont {Sigrist},\ and\ \citenamefont
  {Yanase}}]{yoshida2013}%
  \BibitemOpen
  \bibfield  {author} {\bibinfo {author} {\bibfnamefont {T.}~\bibnamefont
  {Yoshida}}, \bibinfo {author} {\bibfnamefont {M.}~\bibnamefont {Sigrist}}, \
  and\ \bibinfo {author} {\bibfnamefont {Y.}~\bibnamefont {Yanase}},\
  }\href@noop {} {\bibfield  {journal} {\bibinfo  {journal} {Journal of the
  Physical Society of Japan}\ }\textbf {\bibinfo {volume} {82}},\ \bibinfo
  {pages} {074714} (\bibinfo {year} {2013})}\BibitemShut {NoStop}%
\bibitem [{\citenamefont {Houzet}\ \emph {et~al.}(2002)\citenamefont {Houzet},
  \citenamefont {Buzdin}, \citenamefont {Bulaevskii},\ and\ \citenamefont
  {Maley}}]{houzet2002}%
  \BibitemOpen
  \bibfield  {author} {\bibinfo {author} {\bibfnamefont {M.}~\bibnamefont
  {Houzet}}, \bibinfo {author} {\bibfnamefont {A.}~\bibnamefont {Buzdin}},
  \bibinfo {author} {\bibfnamefont {L.}~\bibnamefont {Bulaevskii}}, \ and\
  \bibinfo {author} {\bibfnamefont {M.}~\bibnamefont {Maley}},\ }\href@noop {}
  {\bibfield  {journal} {\bibinfo  {journal} {Physical review letters}\
  }\textbf {\bibinfo {volume} {88}},\ \bibinfo {pages} {227001} (\bibinfo
  {year} {2002})}\BibitemShut {NoStop}%
\bibitem [{\citenamefont {Yoshida}\ \emph {et~al.}(2012)\citenamefont
  {Yoshida}, \citenamefont {Sigrist},\ and\ \citenamefont
  {Yanase}}]{yoshida2012}%
  \BibitemOpen
  \bibfield  {author} {\bibinfo {author} {\bibfnamefont {T.}~\bibnamefont
  {Yoshida}}, \bibinfo {author} {\bibfnamefont {M.}~\bibnamefont {Sigrist}}, \
  and\ \bibinfo {author} {\bibfnamefont {Y.}~\bibnamefont {Yanase}},\
  }\href@noop {} {\bibfield  {journal} {\bibinfo  {journal} {Physical Review
  B}\ }\textbf {\bibinfo {volume} {86}},\ \bibinfo {pages} {134514} (\bibinfo
  {year} {2012})}\BibitemShut {NoStop}%
\bibitem [{\citenamefont {Lu}\ \emph {et~al.}(2015)\citenamefont {Lu},
  \citenamefont {Zheliuk}, \citenamefont {Leermakers}, \citenamefont {Yuan},
  \citenamefont {Zeitler}, \citenamefont {Law},\ and\ \citenamefont
  {Ye}}]{lu2015}%
  \BibitemOpen
  \bibfield  {author} {\bibinfo {author} {\bibfnamefont {J.}~\bibnamefont
  {Lu}}, \bibinfo {author} {\bibfnamefont {O.}~\bibnamefont {Zheliuk}},
  \bibinfo {author} {\bibfnamefont {I.}~\bibnamefont {Leermakers}}, \bibinfo
  {author} {\bibfnamefont {N.~F.}\ \bibnamefont {Yuan}}, \bibinfo {author}
  {\bibfnamefont {U.}~\bibnamefont {Zeitler}}, \bibinfo {author} {\bibfnamefont
  {K.~T.}\ \bibnamefont {Law}}, \ and\ \bibinfo {author} {\bibfnamefont
  {J.}~\bibnamefont {Ye}},\ }\href@noop {} {\bibfield  {journal} {\bibinfo
  {journal} {Science}\ }\textbf {\bibinfo {volume} {350}},\ \bibinfo {pages}
  {1353} (\bibinfo {year} {2015})}\BibitemShut {NoStop}%
\bibitem [{\citenamefont {Xi}\ \emph {et~al.}(2015)\citenamefont {Xi},
  \citenamefont {Wang}, \citenamefont {Zhao}, \citenamefont {Park},
  \citenamefont {Law}, \citenamefont {Berger}, \citenamefont {Forr{\'o}},
  \citenamefont {Shan},\ and\ \citenamefont {Mak}}]{xi2015}%
  \BibitemOpen
  \bibfield  {author} {\bibinfo {author} {\bibfnamefont {X.}~\bibnamefont
  {Xi}}, \bibinfo {author} {\bibfnamefont {Z.}~\bibnamefont {Wang}}, \bibinfo
  {author} {\bibfnamefont {W.}~\bibnamefont {Zhao}}, \bibinfo {author}
  {\bibfnamefont {J.-H.}\ \bibnamefont {Park}}, \bibinfo {author}
  {\bibfnamefont {K.~T.}\ \bibnamefont {Law}}, \bibinfo {author} {\bibfnamefont
  {H.}~\bibnamefont {Berger}}, \bibinfo {author} {\bibfnamefont
  {L.}~\bibnamefont {Forr{\'o}}}, \bibinfo {author} {\bibfnamefont
  {J.}~\bibnamefont {Shan}}, \ and\ \bibinfo {author} {\bibfnamefont {K.~F.}\
  \bibnamefont {Mak}},\ }\href@noop {} {\bibfield  {journal} {\bibinfo
  {journal} {Nature Physics}\ } (\bibinfo {year} {2015})}\BibitemShut {NoStop}%
\bibitem [{\citenamefont {Saito}\ \emph {et~al.}(2016)\citenamefont {Saito},
  \citenamefont {Nakamura}, \citenamefont {Bahramy}, \citenamefont {Kohama},
  \citenamefont {Ye}, \citenamefont {Kasahara}, \citenamefont {Nakagawa},
  \citenamefont {Onga}, \citenamefont {Tokunaga}, \citenamefont {Nojima} \emph
  {et~al.}}]{saito2016}%
  \BibitemOpen
  \bibfield  {author} {\bibinfo {author} {\bibfnamefont {Y.}~\bibnamefont
  {Saito}}, \bibinfo {author} {\bibfnamefont {Y.}~\bibnamefont {Nakamura}},
  \bibinfo {author} {\bibfnamefont {M.~S.}\ \bibnamefont {Bahramy}}, \bibinfo
  {author} {\bibfnamefont {Y.}~\bibnamefont {Kohama}}, \bibinfo {author}
  {\bibfnamefont {J.}~\bibnamefont {Ye}}, \bibinfo {author} {\bibfnamefont
  {Y.}~\bibnamefont {Kasahara}}, \bibinfo {author} {\bibfnamefont
  {Y.}~\bibnamefont {Nakagawa}}, \bibinfo {author} {\bibfnamefont
  {M.}~\bibnamefont {Onga}}, \bibinfo {author} {\bibfnamefont {M.}~\bibnamefont
  {Tokunaga}}, \bibinfo {author} {\bibfnamefont {T.}~\bibnamefont {Nojima}},
  \emph {et~al.},\ }\href@noop {} {\bibfield  {journal} {\bibinfo  {journal}
  {Nature Physics}\ }\textbf {\bibinfo {volume} {12}},\ \bibinfo {pages} {144}
  (\bibinfo {year} {2016})}\BibitemShut {NoStop}%
\bibitem [{\citenamefont {Navarro-Moratalla}\ \emph {et~al.}(2016)\citenamefont
  {Navarro-Moratalla}, \citenamefont {Island}, \citenamefont
  {Ma{\~n}as-Valero}, \citenamefont {Pinilla-Cienfuegos}, \citenamefont
  {Castellanos-Gomez}, \citenamefont {Quereda}, \citenamefont
  {Rubio-Bollinger}, \citenamefont {Chirolli}, \citenamefont
  {Silva-Guill{\'e}n}, \citenamefont {Agra{\"\i}t} \emph
  {et~al.}}]{navarro2016}%
  \BibitemOpen
  \bibfield  {author} {\bibinfo {author} {\bibfnamefont {E.}~\bibnamefont
  {Navarro-Moratalla}}, \bibinfo {author} {\bibfnamefont {J.~O.}\ \bibnamefont
  {Island}}, \bibinfo {author} {\bibfnamefont {S.}~\bibnamefont
  {Ma{\~n}as-Valero}}, \bibinfo {author} {\bibfnamefont {E.}~\bibnamefont
  {Pinilla-Cienfuegos}}, \bibinfo {author} {\bibfnamefont {A.}~\bibnamefont
  {Castellanos-Gomez}}, \bibinfo {author} {\bibfnamefont {J.}~\bibnamefont
  {Quereda}}, \bibinfo {author} {\bibfnamefont {G.}~\bibnamefont
  {Rubio-Bollinger}}, \bibinfo {author} {\bibfnamefont {L.}~\bibnamefont
  {Chirolli}}, \bibinfo {author} {\bibfnamefont {J.~A.}\ \bibnamefont
  {Silva-Guill{\'e}n}}, \bibinfo {author} {\bibfnamefont {N.}~\bibnamefont
  {Agra{\"\i}t}},  \emph {et~al.},\ }\href@noop {} {\bibfield  {journal}
  {\bibinfo  {journal} {Nature communications}\ }\textbf {\bibinfo {volume}
  {7}} (\bibinfo {year} {2016})}\BibitemShut {NoStop}%
\bibitem [{\citenamefont {Zhou}\ \emph {et~al.}(2016)\citenamefont {Zhou},
  \citenamefont {Yuan}, \citenamefont {Jiang},\ and\ \citenamefont
  {Law}}]{zhou2016a}%
  \BibitemOpen
  \bibfield  {author} {\bibinfo {author} {\bibfnamefont {B.~T.}\ \bibnamefont
  {Zhou}}, \bibinfo {author} {\bibfnamefont {N.~F.}\ \bibnamefont {Yuan}},
  \bibinfo {author} {\bibfnamefont {H.-L.}\ \bibnamefont {Jiang}}, \ and\
  \bibinfo {author} {\bibfnamefont {K.~T.}\ \bibnamefont {Law}},\ }\href@noop
  {} {\bibfield  {journal} {\bibinfo  {journal} {Physical Review B}\ }\textbf
  {\bibinfo {volume} {93}},\ \bibinfo {pages} {180501} (\bibinfo {year}
  {2016})}\BibitemShut {NoStop}%
\bibitem [{\citenamefont {Yuan}\ \emph {et~al.}(2014)\citenamefont {Yuan},
  \citenamefont {Mak},\ and\ \citenamefont {Law}}]{yuan2014}%
  \BibitemOpen
  \bibfield  {author} {\bibinfo {author} {\bibfnamefont {N.~F.}\ \bibnamefont
  {Yuan}}, \bibinfo {author} {\bibfnamefont {K.~F.}\ \bibnamefont {Mak}}, \
  and\ \bibinfo {author} {\bibfnamefont {K.~T.}\ \bibnamefont {Law}},\
  }\href@noop {} {\bibfield  {journal} {\bibinfo  {journal} {Physical review
  letters}\ }\textbf {\bibinfo {volume} {113}},\ \bibinfo {pages} {097001}
  (\bibinfo {year} {2014})}\BibitemShut {NoStop}%
\bibitem [{\citenamefont {Frigeri}\ \emph {et~al.}(2004)\citenamefont
  {Frigeri}, \citenamefont {Agterberg}, \citenamefont {Koga},\ and\
  \citenamefont {Sigrist}}]{frigeri2004}%
  \BibitemOpen
  \bibfield  {author} {\bibinfo {author} {\bibfnamefont {P.}~\bibnamefont
  {Frigeri}}, \bibinfo {author} {\bibfnamefont {D.}~\bibnamefont {Agterberg}},
  \bibinfo {author} {\bibfnamefont {A.}~\bibnamefont {Koga}}, \ and\ \bibinfo
  {author} {\bibfnamefont {M.}~\bibnamefont {Sigrist}},\ }\href@noop {}
  {\bibfield  {journal} {\bibinfo  {journal} {Physical review letters}\
  }\textbf {\bibinfo {volume} {92}},\ \bibinfo {pages} {097001} (\bibinfo
  {year} {2004})}\BibitemShut {NoStop}%
\bibitem [{\citenamefont {Zhang}\ \emph {et~al.}(2014)\citenamefont {Zhang},
  \citenamefont {Liu}, \citenamefont {Luo}, \citenamefont {Freeman},\ and\
  \citenamefont {Zunger}}]{zhang2014a}%
  \BibitemOpen
  \bibfield  {author} {\bibinfo {author} {\bibfnamefont {X.}~\bibnamefont
  {Zhang}}, \bibinfo {author} {\bibfnamefont {Q.}~\bibnamefont {Liu}}, \bibinfo
  {author} {\bibfnamefont {J.-W.}\ \bibnamefont {Luo}}, \bibinfo {author}
  {\bibfnamefont {A.~J.}\ \bibnamefont {Freeman}}, \ and\ \bibinfo {author}
  {\bibfnamefont {A.}~\bibnamefont {Zunger}},\ }\href@noop {} {\bibfield
  {journal} {\bibinfo  {journal} {Nature Physics}\ }\textbf {\bibinfo {volume}
  {10}},\ \bibinfo {pages} {387} (\bibinfo {year} {2014})}\BibitemShut
  {NoStop}%
\bibitem [{\citenamefont {Riley}\ \emph {et~al.}(2014)\citenamefont {Riley},
  \citenamefont {Mazzola}, \citenamefont {Dendzik}, \citenamefont {Michiardi},
  \citenamefont {Takayama}, \citenamefont {Bawden}, \citenamefont
  {Graner{\o}d}, \citenamefont {Leandersson}, \citenamefont {Balasubramanian},
  \citenamefont {Hoesch} \emph {et~al.}}]{riley2014}%
  \BibitemOpen
  \bibfield  {author} {\bibinfo {author} {\bibfnamefont {J.~M.}\ \bibnamefont
  {Riley}}, \bibinfo {author} {\bibfnamefont {F.}~\bibnamefont {Mazzola}},
  \bibinfo {author} {\bibfnamefont {M.}~\bibnamefont {Dendzik}}, \bibinfo
  {author} {\bibfnamefont {M.}~\bibnamefont {Michiardi}}, \bibinfo {author}
  {\bibfnamefont {T.}~\bibnamefont {Takayama}}, \bibinfo {author}
  {\bibfnamefont {L.}~\bibnamefont {Bawden}}, \bibinfo {author} {\bibfnamefont
  {C.}~\bibnamefont {Graner{\o}d}}, \bibinfo {author} {\bibfnamefont
  {M.}~\bibnamefont {Leandersson}}, \bibinfo {author} {\bibfnamefont
  {T.}~\bibnamefont {Balasubramanian}}, \bibinfo {author} {\bibfnamefont
  {M.}~\bibnamefont {Hoesch}},  \emph {et~al.},\ }\href@noop {} {\bibfield
  {journal} {\bibinfo  {journal} {Nature Physics}\ }\textbf {\bibinfo {volume}
  {10}},\ \bibinfo {pages} {835} (\bibinfo {year} {2014})}\BibitemShut
  {NoStop}%
\bibitem [{\citenamefont {Dong}\ \emph {et~al.}(2015)\citenamefont {Dong},
  \citenamefont {Wang}, \citenamefont {Zhang}, \citenamefont {Duan},
  \citenamefont {Zhu}, \citenamefont {Sofo},\ and\ \citenamefont
  {Liu}}]{dong2015}%
  \BibitemOpen
  \bibfield  {author} {\bibinfo {author} {\bibfnamefont {X.-Y.}\ \bibnamefont
  {Dong}}, \bibinfo {author} {\bibfnamefont {J.-F.}\ \bibnamefont {Wang}},
  \bibinfo {author} {\bibfnamefont {R.-X.}\ \bibnamefont {Zhang}}, \bibinfo
  {author} {\bibfnamefont {W.-H.}\ \bibnamefont {Duan}}, \bibinfo {author}
  {\bibfnamefont {B.-F.}\ \bibnamefont {Zhu}}, \bibinfo {author} {\bibfnamefont
  {J.~O.}\ \bibnamefont {Sofo}}, \ and\ \bibinfo {author} {\bibfnamefont
  {C.-X.}\ \bibnamefont {Liu}},\ }\href@noop {} {\bibfield  {journal} {\bibinfo
   {journal} {Nature communications}\ }\textbf {\bibinfo {volume} {6}}
  (\bibinfo {year} {2015})}\BibitemShut {NoStop}%
\bibitem [{\citenamefont {Jones}\ \emph {et~al.}(2014)\citenamefont {Jones},
  \citenamefont {Yu}, \citenamefont {Ross}, \citenamefont {Klement},
  \citenamefont {Ghimire}, \citenamefont {Yan}, \citenamefont {Mandrus},
  \citenamefont {Yao},\ and\ \citenamefont {Xu}}]{jones2014}%
  \BibitemOpen
  \bibfield  {author} {\bibinfo {author} {\bibfnamefont {A.~M.}\ \bibnamefont
  {Jones}}, \bibinfo {author} {\bibfnamefont {H.}~\bibnamefont {Yu}}, \bibinfo
  {author} {\bibfnamefont {J.~S.}\ \bibnamefont {Ross}}, \bibinfo {author}
  {\bibfnamefont {P.}~\bibnamefont {Klement}}, \bibinfo {author} {\bibfnamefont
  {N.~J.}\ \bibnamefont {Ghimire}}, \bibinfo {author} {\bibfnamefont
  {J.}~\bibnamefont {Yan}}, \bibinfo {author} {\bibfnamefont {D.~G.}\
  \bibnamefont {Mandrus}}, \bibinfo {author} {\bibfnamefont {W.}~\bibnamefont
  {Yao}}, \ and\ \bibinfo {author} {\bibfnamefont {X.}~\bibnamefont {Xu}},\
  }\href@noop {} {\bibfield  {journal} {\bibinfo  {journal} {Nature Physics}\
  }\textbf {\bibinfo {volume} {10}},\ \bibinfo {pages} {130} (\bibinfo {year}
  {2014})}\BibitemShut {NoStop}%
\bibitem [{\citenamefont {Liu}\ \emph {et~al.}(2013)\citenamefont {Liu},
  \citenamefont {Guo},\ and\ \citenamefont {Freeman}}]{liu2013a}%
  \BibitemOpen
  \bibfield  {author} {\bibinfo {author} {\bibfnamefont {Q.}~\bibnamefont
  {Liu}}, \bibinfo {author} {\bibfnamefont {Y.}~\bibnamefont {Guo}}, \ and\
  \bibinfo {author} {\bibfnamefont {A.~J.}\ \bibnamefont {Freeman}},\
  }\href@noop {} {\bibfield  {journal} {\bibinfo  {journal} {Nano letters}\
  }\textbf {\bibinfo {volume} {13}},\ \bibinfo {pages} {5264} (\bibinfo {year}
  {2013})}\BibitemShut {NoStop}%
\bibitem [{\citenamefont {Sigrist}\ \emph {et~al.}(2014)\citenamefont
  {Sigrist}, \citenamefont {Agterberg}, \citenamefont {Fischer}, \citenamefont
  {Goryo}, \citenamefont {Loder}, \citenamefont {Rhim}, \citenamefont
  {Maruyama}, \citenamefont {Yanase}, \citenamefont {Yoshida},\ and\
  \citenamefont {Youn}}]{sigrist2014}%
  \BibitemOpen
  \bibfield  {author} {\bibinfo {author} {\bibfnamefont {M.}~\bibnamefont
  {Sigrist}}, \bibinfo {author} {\bibfnamefont {D.~F.}\ \bibnamefont
  {Agterberg}}, \bibinfo {author} {\bibfnamefont {M.~H.}\ \bibnamefont
  {Fischer}}, \bibinfo {author} {\bibfnamefont {J.}~\bibnamefont {Goryo}},
  \bibinfo {author} {\bibfnamefont {F.}~\bibnamefont {Loder}}, \bibinfo
  {author} {\bibfnamefont {S.-H.}\ \bibnamefont {Rhim}}, \bibinfo {author}
  {\bibfnamefont {D.}~\bibnamefont {Maruyama}}, \bibinfo {author}
  {\bibfnamefont {Y.}~\bibnamefont {Yanase}}, \bibinfo {author} {\bibfnamefont
  {T.}~\bibnamefont {Yoshida}}, \ and\ \bibinfo {author} {\bibfnamefont
  {S.~J.}\ \bibnamefont {Youn}},\ }\href@noop {} {\bibfield  {journal}
  {\bibinfo  {journal} {Journal of the Physical Society of Japan}\ }\textbf
  {\bibinfo {volume} {83}},\ \bibinfo {pages} {061014} (\bibinfo {year}
  {2014})}\BibitemShut {NoStop}%
\bibitem [{\citenamefont {Goryo}\ \emph {et~al.}(2012)\citenamefont {Goryo},
  \citenamefont {Fischer},\ and\ \citenamefont {Sigrist}}]{goryo2012}%
  \BibitemOpen
  \bibfield  {author} {\bibinfo {author} {\bibfnamefont {J.}~\bibnamefont
  {Goryo}}, \bibinfo {author} {\bibfnamefont {M.~H.}\ \bibnamefont {Fischer}},
  \ and\ \bibinfo {author} {\bibfnamefont {M.}~\bibnamefont {Sigrist}},\
  }\href@noop {} {\bibfield  {journal} {\bibinfo  {journal} {Physical Review
  B}\ }\textbf {\bibinfo {volume} {86}},\ \bibinfo {pages} {100507} (\bibinfo
  {year} {2012})}\BibitemShut {NoStop}%
\bibitem [{\citenamefont {Fischer}\ \emph {et~al.}(2011)\citenamefont
  {Fischer}, \citenamefont {Loder},\ and\ \citenamefont
  {Sigrist}}]{fischer2011}%
  \BibitemOpen
  \bibfield  {author} {\bibinfo {author} {\bibfnamefont {M.~H.}\ \bibnamefont
  {Fischer}}, \bibinfo {author} {\bibfnamefont {F.}~\bibnamefont {Loder}}, \
  and\ \bibinfo {author} {\bibfnamefont {M.}~\bibnamefont {Sigrist}},\
  }\href@noop {} {\bibfield  {journal} {\bibinfo  {journal} {Physical Review
  B}\ }\textbf {\bibinfo {volume} {84}},\ \bibinfo {pages} {184533} (\bibinfo
  {year} {2011})}\BibitemShut {NoStop}%
\bibitem [{\citenamefont {Youn}\ \emph {et~al.}(2012)\citenamefont {Youn},
  \citenamefont {Fischer}, \citenamefont {Rhim}, \citenamefont {Sigrist},\ and\
  \citenamefont {Agterberg}}]{youn2012}%
  \BibitemOpen
  \bibfield  {author} {\bibinfo {author} {\bibfnamefont {S.~J.}\ \bibnamefont
  {Youn}}, \bibinfo {author} {\bibfnamefont {M.~H.}\ \bibnamefont {Fischer}},
  \bibinfo {author} {\bibfnamefont {S.}~\bibnamefont {Rhim}}, \bibinfo {author}
  {\bibfnamefont {M.}~\bibnamefont {Sigrist}}, \ and\ \bibinfo {author}
  {\bibfnamefont {D.~F.}\ \bibnamefont {Agterberg}},\ }\href@noop {} {\bibfield
   {journal} {\bibinfo  {journal} {Physical Review B}\ }\textbf {\bibinfo
  {volume} {85}},\ \bibinfo {pages} {220505} (\bibinfo {year}
  {2012})}\BibitemShut {NoStop}%
\bibitem [{\citenamefont {Nakosai}\ \emph {et~al.}(2012)\citenamefont
  {Nakosai}, \citenamefont {Tanaka},\ and\ \citenamefont
  {Nagaosa}}]{nakosai2012}%
  \BibitemOpen
  \bibfield  {author} {\bibinfo {author} {\bibfnamefont {S.}~\bibnamefont
  {Nakosai}}, \bibinfo {author} {\bibfnamefont {Y.}~\bibnamefont {Tanaka}}, \
  and\ \bibinfo {author} {\bibfnamefont {N.}~\bibnamefont {Nagaosa}},\
  }\href@noop {} {\bibfield  {journal} {\bibinfo  {journal} {Physical review
  letters}\ }\textbf {\bibinfo {volume} {108}},\ \bibinfo {pages} {147003}
  (\bibinfo {year} {2012})}\BibitemShut {NoStop}%
\bibitem [{\citenamefont {Larkin}\ and\ \citenamefont
  {Ovchinnikov}(1965)}]{larkin1965}%
  \BibitemOpen
  \bibfield  {author} {\bibinfo {author} {\bibfnamefont {A.}~\bibnamefont
  {Larkin}}\ and\ \bibinfo {author} {\bibfnamefont {I.}~\bibnamefont
  {Ovchinnikov}},\ }\href@noop {} {\bibfield  {journal} {\bibinfo  {journal}
  {Soviet Physics-JETP}\ }\textbf {\bibinfo {volume} {20}},\ \bibinfo {pages}
  {762} (\bibinfo {year} {1965})}\BibitemShut {NoStop}%
\bibitem [{\citenamefont {Fulde}\ and\ \citenamefont
  {Ferrell}(1964)}]{fulde1964}%
  \BibitemOpen
  \bibfield  {author} {\bibinfo {author} {\bibfnamefont {P.}~\bibnamefont
  {Fulde}}\ and\ \bibinfo {author} {\bibfnamefont {R.~A.}\ \bibnamefont
  {Ferrell}},\ }\href@noop {} {\bibfield  {journal} {\bibinfo  {journal}
  {Physical Review}\ }\textbf {\bibinfo {volume} {135}},\ \bibinfo {pages}
  {A550} (\bibinfo {year} {1964})}\BibitemShut {NoStop}%
\bibitem [{\citenamefont {Xiao}\ \emph {et~al.}(2012)\citenamefont {Xiao},
  \citenamefont {Liu}, \citenamefont {Feng}, \citenamefont {Xu},\ and\
  \citenamefont {Yao}}]{xiao2012}%
  \BibitemOpen
  \bibfield  {author} {\bibinfo {author} {\bibfnamefont {D.}~\bibnamefont
  {Xiao}}, \bibinfo {author} {\bibfnamefont {G.-B.}\ \bibnamefont {Liu}},
  \bibinfo {author} {\bibfnamefont {W.}~\bibnamefont {Feng}}, \bibinfo {author}
  {\bibfnamefont {X.}~\bibnamefont {Xu}}, \ and\ \bibinfo {author}
  {\bibfnamefont {W.}~\bibnamefont {Yao}},\ }\href@noop {} {\bibfield
  {journal} {\bibinfo  {journal} {Physical Review Letters}\ }\textbf {\bibinfo
  {volume} {108}},\ \bibinfo {pages} {196802} (\bibinfo {year}
  {2012})}\BibitemShut {NoStop}%
\bibitem [{\citenamefont {Fu}\ and\ \citenamefont {Berg}(2010)}]{fu2010}%
  \BibitemOpen
  \bibfield  {author} {\bibinfo {author} {\bibfnamefont {L.}~\bibnamefont
  {Fu}}\ and\ \bibinfo {author} {\bibfnamefont {E.}~\bibnamefont {Berg}},\
  }\href@noop {} {\bibfield  {journal} {\bibinfo  {journal} {Physical review
  letters}\ }\textbf {\bibinfo {volume} {105}},\ \bibinfo {pages} {097001}
  (\bibinfo {year} {2010})}\BibitemShut {NoStop}%
\bibitem [{\citenamefont {Fu}(2014)}]{fu2014}%
  \BibitemOpen
  \bibfield  {author} {\bibinfo {author} {\bibfnamefont {L.}~\bibnamefont
  {Fu}},\ }\href@noop {} {\bibfield  {journal} {\bibinfo  {journal} {Physical
  Review B}\ }\textbf {\bibinfo {volume} {90}},\ \bibinfo {pages} {100509}
  (\bibinfo {year} {2014})}\BibitemShut {NoStop}%
\bibitem [{\citenamefont {Ueda}\ and\ \citenamefont {Rice}(1985)}]{ueda1985}%
  \BibitemOpen
  \bibfield  {author} {\bibinfo {author} {\bibfnamefont {K.}~\bibnamefont
  {Ueda}}\ and\ \bibinfo {author} {\bibfnamefont {T.}~\bibnamefont {Rice}},\
  }\href@noop {} {\bibfield  {journal} {\bibinfo  {journal} {Physical Review
  B}\ }\textbf {\bibinfo {volume} {31}},\ \bibinfo {pages} {7114} (\bibinfo
  {year} {1985})}\BibitemShut {NoStop}%
\bibitem [{\citenamefont {Casalbuoni}\ and\ \citenamefont
  {Nardulli}(2004)}]{casalbuoni2004}%
  \BibitemOpen
  \bibfield  {author} {\bibinfo {author} {\bibfnamefont {R.}~\bibnamefont
  {Casalbuoni}}\ and\ \bibinfo {author} {\bibfnamefont {G.}~\bibnamefont
  {Nardulli}},\ }\href@noop {} {\bibfield  {journal} {\bibinfo  {journal}
  {Reviews of Modern Physics}\ }\textbf {\bibinfo {volume} {76}},\ \bibinfo
  {pages} {263} (\bibinfo {year} {2004})}\BibitemShut {NoStop}%
\bibitem [{\citenamefont {Houzet}\ and\ \citenamefont
  {Buzdin}(2001)}]{houzet2001}%
  \BibitemOpen
  \bibfield  {author} {\bibinfo {author} {\bibfnamefont {M.}~\bibnamefont
  {Houzet}}\ and\ \bibinfo {author} {\bibfnamefont {A.}~\bibnamefont
  {Buzdin}},\ }\href@noop {} {\bibfield  {journal} {\bibinfo  {journal}
  {Physical Review B}\ }\textbf {\bibinfo {volume} {63}},\ \bibinfo {pages}
  {184521} (\bibinfo {year} {2001})}\BibitemShut {NoStop}%
\bibitem [{\citenamefont {Agterberg}\ and\ \citenamefont
  {Kaur}(2007)}]{agterberg2007}%
  \BibitemOpen
  \bibfield  {author} {\bibinfo {author} {\bibfnamefont {D.}~\bibnamefont
  {Agterberg}}\ and\ \bibinfo {author} {\bibfnamefont {R.}~\bibnamefont
  {Kaur}},\ }\href@noop {} {\bibfield  {journal} {\bibinfo  {journal} {Physical
  Review B}\ }\textbf {\bibinfo {volume} {75}},\ \bibinfo {pages} {064511}
  (\bibinfo {year} {2007})}\BibitemShut {NoStop}%
\bibitem [{\citenamefont {Chen}\ \emph {et~al.}(2004)\citenamefont {Chen},
  \citenamefont {Vafek}, \citenamefont {Yazdani},\ and\ \citenamefont
  {Zhang}}]{chen2004}%
  \BibitemOpen
  \bibfield  {author} {\bibinfo {author} {\bibfnamefont {H.-D.}\ \bibnamefont
  {Chen}}, \bibinfo {author} {\bibfnamefont {O.}~\bibnamefont {Vafek}},
  \bibinfo {author} {\bibfnamefont {A.}~\bibnamefont {Yazdani}}, \ and\
  \bibinfo {author} {\bibfnamefont {S.-C.}\ \bibnamefont {Zhang}},\ }\href@noop
  {} {\bibfield  {journal} {\bibinfo  {journal} {Physical review letters}\
  }\textbf {\bibinfo {volume} {93}},\ \bibinfo {pages} {187002} (\bibinfo
  {year} {2004})}\BibitemShut {NoStop}%
\bibitem [{\citenamefont {Soto-Garrido}\ and\ \citenamefont
  {Fradkin}(2014)}]{soto2014}%
  \BibitemOpen
  \bibfield  {author} {\bibinfo {author} {\bibfnamefont {R.}~\bibnamefont
  {Soto-Garrido}}\ and\ \bibinfo {author} {\bibfnamefont {E.}~\bibnamefont
  {Fradkin}},\ }\href@noop {} {\bibfield  {journal} {\bibinfo  {journal}
  {Physical Review B}\ }\textbf {\bibinfo {volume} {89}},\ \bibinfo {pages}
  {165126} (\bibinfo {year} {2014})}\BibitemShut {NoStop}%
\bibitem [{\citenamefont {Dimitrova}\ and\ \citenamefont
  {Feigel'Man}(2007)}]{dimitrova2007}%
  \BibitemOpen
  \bibfield  {author} {\bibinfo {author} {\bibfnamefont {O.}~\bibnamefont
  {Dimitrova}}\ and\ \bibinfo {author} {\bibfnamefont {M.}~\bibnamefont
  {Feigel'Man}},\ }\href@noop {} {\bibfield  {journal} {\bibinfo  {journal}
  {Physical Review B}\ }\textbf {\bibinfo {volume} {76}},\ \bibinfo {pages}
  {014522} (\bibinfo {year} {2007})}\BibitemShut {NoStop}%
\bibitem [{\citenamefont {Kaur}\ \emph {et~al.}(2005)\citenamefont {Kaur},
  \citenamefont {Agterberg},\ and\ \citenamefont {Sigrist}}]{kaur2005}%
  \BibitemOpen
  \bibfield  {author} {\bibinfo {author} {\bibfnamefont {R.}~\bibnamefont
  {Kaur}}, \bibinfo {author} {\bibfnamefont {D.}~\bibnamefont {Agterberg}}, \
  and\ \bibinfo {author} {\bibfnamefont {M.}~\bibnamefont {Sigrist}},\
  }\href@noop {} {\bibfield  {journal} {\bibinfo  {journal} {Physical review
  letters}\ }\textbf {\bibinfo {volume} {94}},\ \bibinfo {pages} {137002}
  (\bibinfo {year} {2005})}\BibitemShut {NoStop}%
\bibitem [{\citenamefont {Michaeli}\ \emph {et~al.}(2012)\citenamefont
  {Michaeli}, \citenamefont {Potter},\ and\ \citenamefont
  {Lee}}]{michaeli2012}%
  \BibitemOpen
  \bibfield  {author} {\bibinfo {author} {\bibfnamefont {K.}~\bibnamefont
  {Michaeli}}, \bibinfo {author} {\bibfnamefont {A.~C.}\ \bibnamefont
  {Potter}}, \ and\ \bibinfo {author} {\bibfnamefont {P.~A.}\ \bibnamefont
  {Lee}},\ }\href@noop {} {\bibfield  {journal} {\bibinfo  {journal} {Physical
  review letters}\ }\textbf {\bibinfo {volume} {108}},\ \bibinfo {pages}
  {117003} (\bibinfo {year} {2012})}\BibitemShut {NoStop}%
\bibitem [{\citenamefont {Yang}\ and\ \citenamefont
  {Agterberg}(2000)}]{yang2000}%
  \BibitemOpen
  \bibfield  {author} {\bibinfo {author} {\bibfnamefont {K.}~\bibnamefont
  {Yang}}\ and\ \bibinfo {author} {\bibfnamefont {D.}~\bibnamefont
  {Agterberg}},\ }\href@noop {} {\bibfield  {journal} {\bibinfo  {journal}
  {Physical review letters}\ }\textbf {\bibinfo {volume} {84}},\ \bibinfo
  {pages} {4970} (\bibinfo {year} {2000})}\BibitemShut {NoStop}%
\bibitem [{\citenamefont {Efimkin}\ and\ \citenamefont
  {Lozovik}(2011)}]{efimkin2011}%
  \BibitemOpen
  \bibfield  {author} {\bibinfo {author} {\bibfnamefont {D.}~\bibnamefont
  {Efimkin}}\ and\ \bibinfo {author} {\bibfnamefont {Y.~E.}\ \bibnamefont
  {Lozovik}},\ }\href@noop {} {\bibfield  {journal} {\bibinfo  {journal}
  {Journal of Experimental and Theoretical Physics}\ }\textbf {\bibinfo
  {volume} {113}},\ \bibinfo {pages} {880} (\bibinfo {year}
  {2011})}\BibitemShut {NoStop}%
\bibitem [{\citenamefont {Seradjeh}(2012)}]{seradjeh2012}%
  \BibitemOpen
  \bibfield  {author} {\bibinfo {author} {\bibfnamefont {B.}~\bibnamefont
  {Seradjeh}},\ }\href@noop {} {\bibfield  {journal} {\bibinfo  {journal}
  {Physical Review B}\ }\textbf {\bibinfo {volume} {85}},\ \bibinfo {pages}
  {235146} (\bibinfo {year} {2012})}\BibitemShut {NoStop}%
\end{thebibliography}
\end{document}